\documentclass[pdflatex, iicol]{sn-jnl}

\usepackage[T1]{fontenc}
\usepackage[utf8]{inputenc}
\usepackage{graphicx}
\usepackage{epsf}
\usepackage{bm}
\usepackage{amsmath,amssymb,amsfonts}
\usepackage{epstopdf}
\usepackage{rotating}
\usepackage{pdflscape}
\usepackage{adjustbox}
\usepackage{xcolor}
\usepackage{dcolumn}
\usepackage{ulem}
\usepackage{appendix}
\usepackage{enumitem}
\usepackage{booktabs}
\usepackage{multirow}
\usepackage{nicefrac}
\usepackage{hyperref}

\newcommand{\orcidlink}[1]{}

\begin{document}

\title[Self-gravitating logarithmic condensates]{Equilibrium structure and radial dynamics of self-gravitating logarithmic condensates}

\author*[1,2]{\fnm{Jorge} \sur{Mastache}}
\email{jh.mastache@mctp.mx}

\author[3]{\fnm{El\'ias} \sur{Castellanos}}
\email{elias.castellanos@tec.mx}

\author[4]{\fnm{Guillermo} \sur{Chac\'on--Acosta}}
\email{gchacon@cua.uam.mx}

\affil*[1]{\orgname{Secretar\'ia de Ciencia, Humanidades, Tecnolog\'ia e Innovaci\'on},
\orgaddress{\street{Av. Insurgentes Sur 1582, Col. Cr\'edito Constructor, Alc. Benito Ju\'arez},
\postcode{03940}, \city{Ciudad de M\'exico}, \country{M\'exico}}}

\affil[2]{\orgdiv{Mesoamerican Centre for Theoretical Physics},
\orgname{Universidad Aut\'onoma de Chiapas},
\orgaddress{\street{Ciudad Universitaria, Carretera Zapata Km. 4, Real del Bosque (Ter\'an)},
\postcode{29040}, \city{Tuxtla Guti\'errez, Chiapas}, \country{M\'exico}}}

\affil[3]{\orgdiv{Escuela de Ingenier\'ia y Ciencias},
\orgname{Tecnol\'ogico de Monterrey},
\orgaddress{\street{Carr. Lago de Guadalupe Km. 3.5},
\postcode{52926}, \city{Estado de M\'exico}, \country{M\'exico}}}

\affil[4]{\orgdiv{Departamento de Matem\'aticas Aplicadas y Sistemas},
\orgname{Universidad Aut\'onoma Metropolitana-Cuajimalpa},
\orgaddress{\street{Av. Vasco de Quiroga 4871},
\postcode{05348}, \city{Ciudad de M\'exico}, \country{M\'exico}}}

\abstract{ 
Contact and logarithmic interactions can support a variety of self-gravitating condensates, with different equilibrium sizes and responses to radial perturbations. We study these configurations in the Newtonian regime using a time-dependent Gaussian approximation. Our main aim is to understand how the signs and relative strengths of the interactions organize the equilibrium parameter space. Repulsive contact interactions admit a unique positive Gaussian equilibrium when the logarithmic coupling is nonnegative. For positive logarithmic coupling, four limiting balances combine gradient or contact support with gravitational or logarithmic confinement. Transitions between these balances preserve radial stability, whereas folds in the other sign sectors mark the merger of equilibrium branches. When both couplings are negative, stability requires gravity to dominate logarithmic expansion. The response of the equilibrium radius to the logarithmic coupling provides a direct measure of these changes. We explore the resulting parameter space through mass--radius benchmarks at solar, dwarf-galaxy, and galaxy-cluster scales. The numerical selection uses compactness and diluteness criteria, with additional energy-scale checks at representative configurations. The same effective potential determines radial oscillations, escape barriers, and the pressure associated with collective compression. Comparing four trial profiles reveals how the predicted radii and oscillation frequencies depend on the assumed density distribution. These results connect the structure of the parameter maps to the forces governing confinement and radial stability in logarithmic bosonic condensates.
}

\keywords{boson clouds, logarithmic interactions, Bose-Einstein condensate, boson stars, dark matter}

\maketitle

\section{Introduction} \label{sec:intro}

Self-gravitating bosonic configurations provide a setting in which quantum effects, interparticle interactions, and gravity jointly determine the structure and dynamics of a macroscopic system. Relativistic equilibrium configurations of self-gravitating massive scalar fields provide the basis for boson-star models \cite{Kaup,Ruffini}. Repulsive scalar self-interactions introduce an additional support mechanism and can substantially increase the characteristic mass scale of equilibrium boson stars \cite{Colpi:1986ye}. Subsequent work has developed a broad range of boson-star models, characterized by different self-interactions, compactnesses, and dynamical properties \cite{visinelli,Liebling:2023review}. This diversity motivates a systematic investigation of how specific interactions determine equilibrium structure and collective stability.

At galactic scales, the possibility that dark matter possesses a coherent bosonic description has motivated models of halos as Bose liquids or self-gravitating condensates \cite{sin,Boehmer:2007BEC}. The wave properties of ultralight bosons provide a characteristic scale for gravitational structure, linking particle mass to the suppression of small-scale clustering \cite{Hu:2000ke}. Cosmological simulations of wave dark matter show self-gravitating solitonic cores embedded in extended fluctuating halos, linking coherent equilibrium structures to halo assembly \cite{Schive:2014dra,Schive:2014hza}. The connection between condensate models and galactic observables has been explored through rotation-curve analyses of self-interacting scalar clouds in a central gravitational background, including comparisons with standard halo profiles \cite{Castellanos:2019ttq,Castellanos:2023qso}. These studies motivate an examination of how microscopic interactions determine the size and density structure of bosonic configurations across astrophysical scales.

In the nonrelativistic regime, the mean-field dynamics of a dilute condensate with short-range interactions are described by the Gross--Pitaevskii equation \cite{Gross:1961kqh,Pitaevskii:1961vortex}. When coupled to the Newtonian Poisson equation, it incorporates the gravitational field generated by the condensate density. The resulting balance between gradient energy, contact interactions, and gravitational attraction determines characteristic mass--radius relations \cite{Chavanis:2011zi,Chavanis:2011zm}. Stationary solutions of the Gross--Pitaevskii--Poisson system further demonstrate how treating the particle number $N$ as an independent macroscopic control parameter organizes the equilibrium families and their characteristic radii \cite{Guzman:2026slq}. The resulting dependence on $(m_\phi,a,N)$ provides a direct basis for exploring how an additional logarithmic interaction reshapes the equilibrium parameter space. This framework connects microscopic parameters to macroscopic observables and supports effective descriptions of halos with a coherent central core and an outer atmosphere \cite{Chavanis:2021dmm}. Astrophysical mass and radius scales thus provide useful benchmarks for exploring the mechanisms of confinement and structural properties of condensate configurations.

The equilibrium and dynamical properties of self-gravitating condensates are particularly sensitive to the form and sign of their interactions. Time-dependent studies of relativistic boson stars have characterized stable and unstable branches through their response to finite perturbations \cite{Seidel:1990jh}. For contact interactions, analytical variational methods and numerical solutions have established mass--radius sequences and stability criteria, including the qualitative difference between repulsive and attractive couplings \cite{Chavanis:2011zi,Chavanis:2011zm}. Linear perturbation analyses of the Gross--Pitaevskii--Poisson system establish the stability of ground-state configurations with repulsive contact interactions and identify a maximum-mass configuration separating stable and unstable branches for attractive interactions \cite{Nambo:2024gvs}. Time-dependent variational methods connect condensate widths and phases to collective excitation frequencies and nonlinear evolution \cite{PerezGarcia:1996variational,PerezGarcia:1997dynamics}. Their application to self-gravitating systems provides a useful description of contraction and collapse in attractive configurations \cite{Chavanis:2016dab}. Their practical value lies in reducing an infinite-dimensional field problem to a small set of mechanical variables. Comparisons between different ansatz families and numerical solutions quantify how sensitive physical observables are to the trial profile \cite{Eby:2018dat}. An extension of the interaction energy therefore calls for a corresponding examination of both equilibrium stability and profile dependence.

A direct motivation comes from self-bound states in ultracold Bose gases. The dilute-gas expansion establishes the leading quantum-fluctuation correction to the mean-field energy \cite{Lee:1957zzb}. In Bose mixtures, the interaction of this correction with competing contact interactions can stabilize a self-bound liquid \cite{Petrov:2015droplets}, and self-bound droplets were subsequently realized experimentally in atomic mixtures \cite{Cabrera:2018droplets,Semeghini}. These developments show how competing energy contributions can support a finite equilibrium size without external confinement. Furthermore, in references\cite{RodriguezLopez:2021JLTP,Castellanosdroplets2:2025omh}, it was demonstrated that a logarithmic potential gives rise to the formation of "quantum droplet"-like configurations in a single-component Bose gas (without the need for Bose mixtures); that is, upon removing the external confining potential, self-sustainability—or self-confinement—emerges. This self-confining behavior also manifests in the relativistic regime, within the framework of a flat geometry of space-time \cite{Hernandez:2026xti}. 

The interplay between self-confinement and collective motion motivates exploring effective interactions that support localized condensate configurations and their extension to self-gravitating systems.

The logarithmic Schr\"odinger equation provides an analytically tractable nonlinear model with localized Gaussian solutions \cite{Bialynicki-Birula:1976tja}. In Bose liquids, the logarithmic term can support self-sustained configurations without an external trapping potential \cite{Avdeenkov:2011mj}. Its dynamical implications have been examined in studies of condensate expansion and prospective experimental bounds \cite{Vowe:2020mqp}, as well as in the emergence of oscillating localized states in one-dimensional logarithmic Bose gases \cite{RodriguezLopez:2021JLTP}. Statistical and hydrodynamic interpretations motivate logarithmic nonlinearities as effective descriptions of interacting quantum fluids \cite{Zloshchastiev:2025pgf}. Relativistic scalar models with cubic and logarithmic interactions also yield a logarithmic Gross--Pitaevskii equation in the nonrelativistic limit \cite{Hernandez:2026xti}. We adopt this effective description to examine how the logarithmic coupling controls confinement and radial dynamics.

Logarithmic interactions have also been directly studied in self-gravitating systems. A generalized Gross--Pitaevskii--Poisson framework includes arbitrary nonlinearity and dissipation, including cubic and logarithmic terms, and yields a Gaussian variational equation for the condensate size \cite{Chavanis:2016ial}. Related work interprets a logarithmic term and damping as an effective description of relaxation toward a core--halo configuration \cite{Chavanis:2021dmm}. In the Thomas--Fermi approximation, logarithmic condensates have been investigated in gravitational-collapse calculations \cite{Stefanescu:2023} and in models of halo density profiles and rotation curves \cite{Haghani:2025aho}. Together, these developments motivate a systematic analysis of how logarithmic interactions reshape the balance between quantum pressure, contact interactions, and self-gravity. Connecting this balance to equilibrium branches and collective radial motion provides a direct way to identify the physical mechanisms governing confinement and stability.

The main application is the organization of the interaction-dependent parameter space across stellar and halo benchmarks. Analytical existence, stability, and fold conditions guide the interpretation of the numerical slices, while explicit validity criteria identify representative configurations within the effective description. The collective pressure response and trial-profile comparison then characterize how equilibrium predictions depend on compression and the assumed density distribution.

We use this framework to organize the microscopic parameter space across stellar and halo scales. Analytical branch conditions and the coupling response identify the mechanisms controlling equilibrium size and radial stability. Benchmark comparisons, collective pressure, and trial-profile variations then connect these mechanisms to the structural properties of the condensate.

This paper is structured as follows: Section~\ref{sec:stability-conditions} develops the field model, Gaussian reduction, radial dynamics, and analytical branch classification. Section~\ref{sec:results-v5} presents the general equilibrium response and the pressure relation before examining the benchmark selections, dwarf-halo maps, and profile sensitivity. Section~\ref{sec:conclusions} summarizes the physical implications. The appendices give the solar and cluster maps and the four-profile comparison.

\section{Self--gravitating variational model} \label{sec:stability-conditions}  

\subsection{Energy functional, field equation, and sign convention}
We consider a nonrelativistic condensate of $N$ bosons of mass $m_\phi$, described by a wave function $\Psi(\mathbf r,t)$ normalized to $\int |\Psi|^2d^3r=N$. The particles interact through a contact term and an effective logarithmic nonlinearity, while the condensate density generates its own Newtonian gravitational field. The conserved energy is
\begin{align}
E[\Psi]={}& \int d^3r\left[ \frac{\hbar^2}{2m_\phi}|\nabla\Psi|^2 +\frac{g}{2}|\Psi|^4 \right] \notag\\
&-\beta\int d^3r\,|\Psi|^2 \left[\ln(\alpha^3|\Psi|^2)-1\right] \notag\\ 
&+\frac12\int d^3r\,\rho_m\Phi.
\label{eq:log-erg-func-3d}
\end{align}
where $g=4\pi\hbar^2a/m_\phi$, $a$ is the scattering length, $\beta$ has units of energy, and $\alpha$ is a reference length. The mass density and gravitational potential satisfy
\begin{align}
 \rho_m&=m_\phi|\Psi|^2, &
 \Phi(\mathbf r)&=-G\int\frac{\rho_m(\mathbf r')}{|\mathbf r-\mathbf r'|}\,d^3r',
 \label{eq:newtonian-potential}
\end{align}
The gravitational potential satisfies Poisson's equation $\nabla^2\Phi=4\pi G\rho_m$, with the boundary condition $\Phi(r)\to0$ as $r\to\infty$. Variation with respect to $\Psi^*$, with $\Phi$ determined by the density, gives
\begin{equation}
 i\hbar\frac{\partial\Psi}{\partial t}
 =\left[-\frac{\hbar^2}{2m_\phi}\nabla^2+g|\Psi|^2
 -\beta\ln(\alpha^3|\Psi|^2)+m_\phi\Phi\right]\Psi.
 \label{eq:log-gpp}
\end{equation}
The factor $1/2$ in the gravitational energy accounts for pair counting; the field equation contains the full potential $m_\phi\Phi$. This formulation combines the usual contact-interaction description with the logarithmic extension of condensate dynamics \cite{Pethick:2008book,Avdeenkov:2011mj,Chavanis:2016ial}.

Our convention assigns a repulsive contact interaction to $g>0$ and an attractive one to $g<0$. The logarithmic contribution favors contraction for $\beta>0$ and expansion for $\beta<0$, as the radial equation below makes explicit. Changing $\alpha$ to $\alpha'$ shifts the energy at fixed $N$ by $-3\beta N\ln(\alpha'/\alpha)$ and changes only a spatially uniform phase in Eq.~\eqref{eq:log-gpp}. Equilibrium forces and oscillation frequencies are therefore independent of this reference length.

We treat $\beta$ as an effective coupling characterizing the density-dependent interaction energy. Its local pressure contribution is $p_{\log}=-\beta n$, where $n=|\Psi|^2$. Thus $\beta<0$ supplies positive pressure linear in density, while $\beta>0$ favors self-confinement. Statistical, fluid, and relativistic constructions provide complementary motivations for this effective description \cite{Avdeenkov:2011mj,Chavanis:2016ial,Zloshchastiev:2025pgf, Hernandez:2026xti}.

\subsection{Gaussian reduction and physical radius}
A Gaussian profile permits an analytical evaluation of each contribution to the energy and a direct connection between equilibrium and collective motion \cite{Chavanis:2011zi,Chavanis:2011zm}. Gaussian width dynamics also describe collective oscillations and stability in harmonically trapped logarithmic condensates \cite{Bouharia2015LogTrap}. We use the normalized time-dependent ansatz
\begin{equation}
 \Psi(r,t)=\frac{N^{1/2}}{\pi^{3/4}R^{3/2}}
 \exp\left(-\frac{r^2}{2R^2}\right)
 \exp\left[i\frac{m_\phi\dot R}{2\hbar R}r^2+i\varphi_0(t)\right],
 \label{eq:gaussian-ansatz}
\end{equation}
where $R=R(t)$. The quadratic phase produces the homologous velocity $\mathbf v=(\dot R/R)\mathbf r$ and satisfies the continuity equation. The central density, potential, and enclosed mass are
\begin{align}
 \rho_c(R)&=\frac{m_\phi N}{\pi^{3/2}R^3},
 \label{eq:central-density}\\
 \Phi(r;R)&=-\frac{Gm_\phi N}{r}\operatorname{erf}(r/R),
 \label{eq:gaussian-potential}\\
 M(<r)&=m_\phi N\left[\operatorname{erf}(r/R)
 -\frac{2r}{\sqrt\pi R}e^{-r^2/R^2}\right].
 \label{eq:enclosed-mass}
\end{align}
The gravitational potential remains finite at the center, where $\Phi(0;R)=-2Gm_\phi N/(\sqrt{\pi}R)$. The total mass is $M=m_\phi N$. A mass-fraction radius is written as $R_f=\kappa_f R$, where
\begin{equation}
 \operatorname{erf}(\kappa_f)-\frac{2\kappa_f}{\sqrt\pi}e^{-\kappa_f^2}=f.
 \label{eq:kappa-definition}
\end{equation}
In particular, $R_{99}=\kappa_{99}R$ with $\kappa_{99}\simeq2.38169$. The width and enclosed-mass radius provide complementary measures of size, whose profile dependence can be assessed by comparing variational families \cite{Eby:2018dat,Castellanos:2020omh}.

The phase contribution to the energy is $T_R=3Nm_\phi\dot R^2/4$. The remaining terms define
\begin{align}
U(R)={}& \frac{3\hbar^2N}{4m_\phi R^2} +\frac{gN^2}{2(2\pi)^{3/2}R^3} - \frac{Gm_\phi^2N^2}{\sqrt{2\pi}R} \notag\\
&-\beta N\left[ \ln\left(\frac{\alpha^3N}{\pi^{3/2}R^3}\right) -\frac52 \right].
\label{eq:effective-potential}
\end{align}
These are the gradient, contact, gravitational, and logarithmic energies. The gravitational term follows from the exact Newtonian self-energy of the Gaussian density. The reduced energy is $E_R=T_R+U(R)$.

\subsection{Characteristic scales and equilibrium}
\label{sec:equilibrium-stability}
The equilibrium width $R_0$ satisfies
\begin{equation}
 -\frac{3N\hbar^2}{2m_\phi R_0^3}
 -\frac{3gN^2}{2(2\pi)^{3/2}R_0^4}
 +\frac{Gm_\phi^2N^2}{\sqrt{2\pi}R_0^2}
 +\frac{3\beta N}{R_0}=0.
 \label{eq:stationarity-dimensional}
\end{equation}
We retain the characteristic scales
\begin{equation}
 \ell_\beta^3 \equiv \frac{gN}{2(2\pi)^{3/2}\beta},\quad
 \ell_G^2 \equiv \frac{3g}{4\pi Gm_\phi^2},\quad
 \ell_Q \equiv \frac{m_\phi gN}{(2\pi)^{3/2}\hbar^2}.
 \label{eq:ells}
\end{equation}
For $g>0$ and $\beta>0$, these define positive lengths. Across sign sectors, $\ell_\beta^3$, $\ell_G^2$, and $\ell_Q$ are used as signed algebraic quantities: in particular, $\ell_G^2<0$ for attractive contact interactions. This convention keeps the signs explicit without taking square roots of negative quantities. At $\beta=0$ we set $1/\ell_\beta^3=0$. The limit $g=0$ is evaluated directly using the dimensional equations.

For $g\ne0$, Eq.~\eqref{eq:stationarity-dimensional} becomes
\begin{equation}
 \frac{R_0^3}{\ell_\beta^3}+\frac{R_0^2}{\ell_G^2}
 -\frac{R_0}{\ell_Q}-1=0.
 \label{eq:equilibrium-length-scales}
\end{equation}
More generally,
\begin{equation}
 U'(R)=\frac{3gN^2}{2(2\pi)^{3/2}R^4}
 \left[\frac{R^3}{\ell_\beta^3}+\frac{R^2}{\ell_G^2}
 -\frac{R}{\ell_Q}-1\right].
 \label{eq:force-factorization}
\end{equation}
This expression relates the equilibrium polynomial to the physical radial force, including the sign of $g$.

\subsection{Domain of the nonrelativistic description}
\label{sec:validity}
For the Gaussian density, weak gravity and diluteness are monitored by
\begin{equation}
 \frac{|\Phi(0)|}{c^2}=\frac{2GM}{\sqrt\pi Rc^2}\ll1,
 \qquad \frac{N|a|^3}{\pi^{3/2}R^3}\ll1.
 \label{eq:gravity-dilution-cuts}
\end{equation}
The nonrelativistic treatment also requires small kinetic and interaction scales relative to the rest energy. Useful checks are
\begin{equation}
 \frac{3\hbar^2}{4m_\phi^2c^2R^2}\ll1,
 \quad \frac{|g|N}{\pi^{3/2}R^3m_\phi c^2}\ll1,
 \quad \frac{|\beta|\kappa_{99}^2}{m_\phi c^2}\ll1.
 \label{eq:nonrelativistic-cuts}
\end{equation}
The first ratio is the gradient energy per particle divided by $m_\phi c^2$; the second bounds the central contact mean-field scale. The third controls the spatial variation of the logarithmic mean-field potential across the region containing $99\%$ of the mass, since
$|\beta\ln[|\Psi(R_{99})|^2/|\Psi(0)|^2]|=|\beta|\kappa_{99}^2$.
It is independent of the reference $\alpha$. During homologous evolution, we additionally monitor
\begin{equation}
  \frac{\kappa_{99}^2\dot R^2}{c^2}\ll1.
  \label{eq:flow-validity}
\end{equation}
These conditions specify the domain of the effective description within the mass-containing region. We evaluate them at equilibrium and, for astrophysical interpretations of time-dependent solutions, along the corresponding trajectories.

\subsection{Radial equation and small oscillations}
\label{sec:collective-dynamics}
The Gaussian phase in Eq.~\eqref{eq:gaussian-ansatz} gives
\begin{equation}
 T_R=\frac{3Nm_\phi}{4}\dot R^2=\frac12M_R\dot R^2,
 \qquad M_R=\frac32Nm_\phi.
 \label{eq:collective-kinetic-energy}
\end{equation}
The reduced Lagrangian $L_R=T_R-U(R)$ yields $M_R\ddot R=-U'(R)$ and hence
\begin{equation}
 \ddot R=\frac{\hbar^2}{m_\phi^2R^3}
 +\frac{gN}{m_\phi(2\pi)^{3/2}R^4}
 -\frac{2Gm_\phi N}{3\sqrt{2\pi}R^2}
 -\frac{2\beta}{m_\phi R}.
 \label{eq:radial-dimensional}
\end{equation}
For $g\ne0$ the same equation is
\begin{equation}
 \ddot R=\frac{gN}{m_\phi(2\pi)^{3/2}R^4}
 \left[1+\frac{R}{\ell_Q}-\frac{R^2}{\ell_G^2}
 -\frac{R^3}{\ell_\beta^3}\right].
 \label{eq:radial-lengths}
\end{equation}
These equations extend the collective-coordinate description of self-gravitating condensates to the present sign convention and conservative setting \cite{Chavanis:2011zi,Chavanis:2016ial}.

For $R(t)=R_0+\delta R(t)$, the linearized equation is
\begin{equation}
 \delta\ddot R+\Omega_R^2\delta R=0,
 \label{eq:radial-frequency-definition}
\end{equation}
with $\Omega_R^2=\frac{U''(R_0)}{M_R}$ being
\begin{align}
 \Omega_R^2=\frac{gN}{m_\phi(2\pi)^{3/2}R_0^5}
 \left[1+\frac{R_0^2}{\ell_G^2}
 +\frac{2R_0^3}{\ell_\beta^3}\right],
 \label{eq:radial-frequency-equilibrium}
\end{align}
in which we used the equilibrium equation. Stable equilibria have $\Omega_R^2>0$, with period $2\pi/\Omega_R$, while $\Omega_R^2<0$ gives exponential growth in the homologous radial mode. At a marginal equilibrium, $\Omega_R^2=0$.

\subsection{Conservative trajectories and energy barriers}
We use $x=R/R_0$ and $y=\Omega_0t$, where $\Omega_0=\Omega_R(R_0)>0$ is the small-oscillation frequency at the selected stable equilibrium. The energy scale is $E_0=M_RR_0^2\Omega_0^2=R_0^2U''(R_0)$. In this normalization,
\begin{align}
\mathcal U(x)
&=\frac{U(R_0x)-U(R_0)}{E_0} \notag\\
&=\frac{3gN^2}{2(2\pi)^{3/2}R_0^3E_0}
\Biggl[
\frac{R_0}{2\ell_Q}(x^{-2}-1)
+\frac{x^{-3}-1}{3} \notag\\
&\qquad
-\frac{R_0^2}{\ell_G^2}(x^{-1}-1)
+\frac{R_0^3}{\ell_\beta^3}\ln x
\Biggr].
\label{eq:dimensionless-potential}
\end{align}
The evolution and conserved energy are
\begin{equation}
x''=-\frac{d\mathcal U}{dx},\qquad
\frac12(x')^2+\mathcal U(x)=\mathcal E,
\label{eq:dimensionless-conservation}
\end{equation}
where primes denote derivatives with respect to $y$. This choice gives $\mathcal U(1)=0$ and $\mathcal U''(1)=1$. For $x(0)=1+\epsilon$ and $x'(0)=0$, the linear limit is $x=1+\epsilon\cos y+O(\epsilon^2)$. Dimensional evolution remains well-defined at marginality, where the normalization by $\Omega_0$ becomes singular.

\begin{figure*}[htbp]
\centering
    \includegraphics[width=\linewidth]{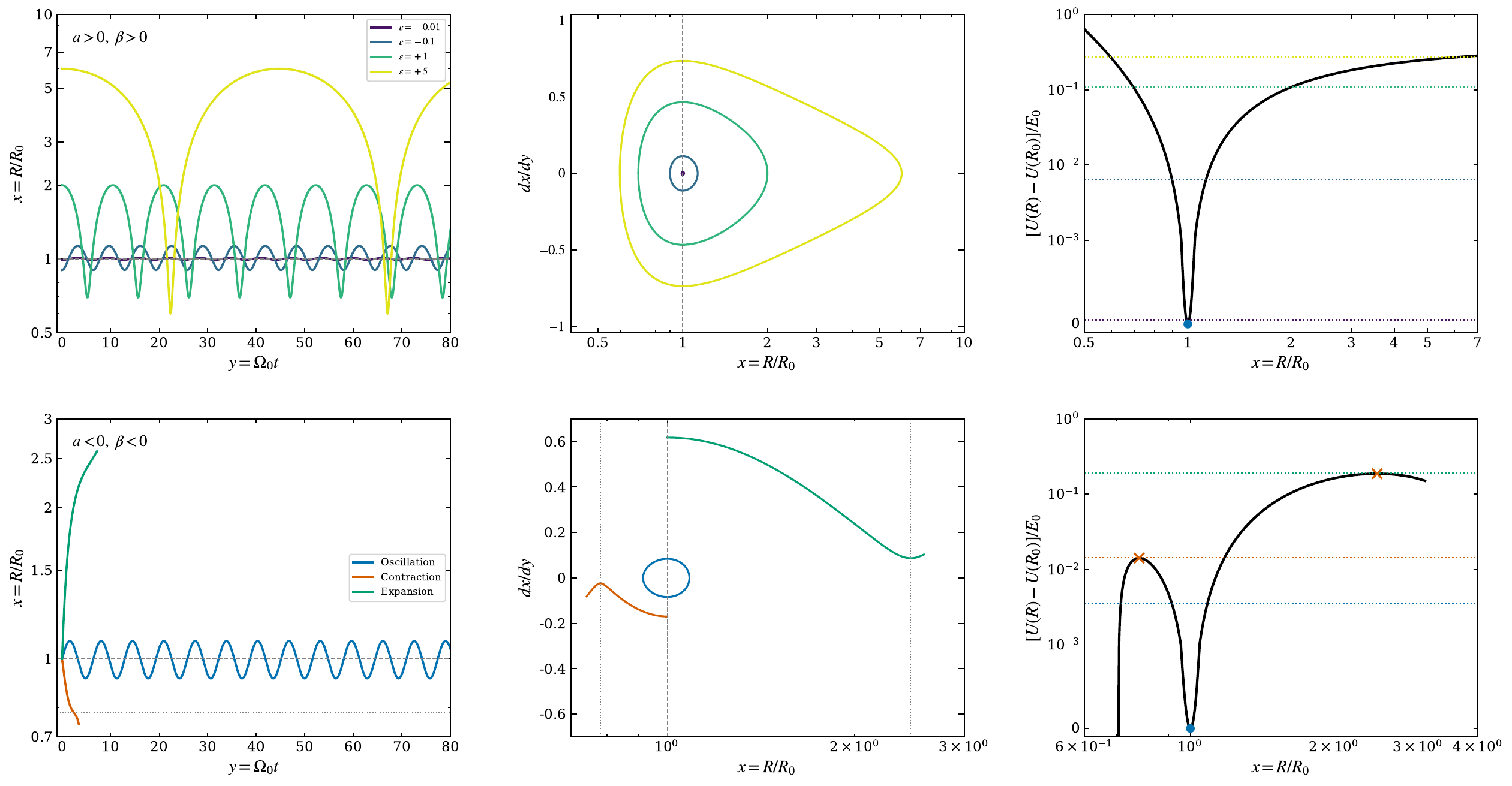}
    \caption{Conservative radial motion in the Gaussian model. Columns show $x(y)$, $(x,x')$, and $\mathcal U(x)$; dotted horizontal lines in the potential panels mark the trajectory energies. The positive-sector trajectories start at $x=1+\epsilon$, $x'=0$, with the indicated displacements. The negative-sector trajectories start at $x=1$: the oscillatory speed is one half of the smaller barrier speed, while contraction and expansion use speeds $1\%$ above their respective thresholds. The latter trajectories terminate at numerical radii just beyond the barriers. The positive reference has $R_0=2.8260\times10^{20}\,{\rm m}$. The negative reference was selected to resolve both barriers and has $R_0=8.0795\times10^{14}\,{\rm m}$; the corresponding plot illustrates the barrier structure.}
    \label{fig:unified-dynamics}
\end{figure*}

Field simulations and variational calculations characterize breathing and quadrupole oscillations in self-gravitating condensates \cite{Asakawa:2023tkm}, while self-consistent Bogoliubov--de Gennes analyses resolve their collective-mode spectrum \cite{Salasnich:2025nrt}. These results provide benchmarks for assessing the frequency represented by our Gaussian radial coordinate. Figure~\ref{fig:unified-dynamics} connects the radius history, phase portrait, and potential landscape. In the positive sector, $a>0$ and $\beta>0$, the two divergent endpoints of $U$ enclose every finite-energy orbit. Larger displacements produce asymmetric turning points and an amplitude-dependent period. In the negative sector, two maxima can bound a local minimum. Energies below both barriers produce closed oscillations; exceeding the inner or outer barrier permits contraction or expansion, respectively. For a trajectory launched from $x=1$, the threshold speed is $|x'(0)|=\sqrt{2\mathcal U(x_{\rm barrier})}$. These mechanisms motivate the root classification below.

The plotted integrations use a maximum step of $0.05$ in $y$ and a maximum duration of $y=80$. The energy diagnostic is $\max|\tfrac12(x')^2+\mathcal U-\mathcal E(0)|/\max(1,|\mathcal E(0)|)$. Reintegrating the archived configurations gives a maximum drift of $1.6\times10^{-12}$. Tightening the tolerances to $10^{-12}$ and $10^{-14}$ and halving the maximum step changes $x(y)$ by less than $2.9\times10^{-10}$ on the common integration intervals, and the barrier-crossing termination times by less than $2.4\times10^{-12}$ in $y$. The negative trajectories end at $0.95$ times the inner-barrier radius or $1.05$ times the outer-barrier radius; these are numerical stopping conditions. The conservation and refinement tests establish the numerical accuracy of the reduced trajectories. Astrophysical applications require configurations satisfying the conditions in Sec.~\ref{sec:validity} throughout their evolution.

\subsection{Uniqueness, stability, and folds}
At a stationary point, the radial curvature is
\begin{align}
 U''(R_0) =\frac{3gN^2}{2(2\pi)^{3/2}R_0^5}
 \left[1+\frac{R_0^2}{\ell_G^2}+\frac{2R_0^3}{\ell_\beta^3}\right].
 \label{eq:stability-length-scales}
\end{align}
A positive curvature gives stability against homologous radial perturbations. For $g<0$, the sign of the prefactor must be included when interpreting the bracket.

For $g>0$ and $\beta\geq0$, the equilibrium polynomial is negative at $R=0$ and positive for sufficiently large $R$. Its nonzero coefficients, ordered by descending power, have exactly one sign change. Descartes' rule therefore gives exactly one positive root, including the quadratic limit $\beta=0$. Equation~\eqref{eq:stability-length-scales} is strictly positive at that root. Thus, this entire sector has a unique radially stable Gaussian equilibrium. This is an analytical result; a parameter scan samples the branch and locates its physical regimes.

A marginal stationary point satisfies $U'(R_{\rm crit})=U''(R_{\rm crit})=0$. Eliminating $1/\ell_Q$ gives
\begin{equation}
 \frac{2R_{\rm crit}^3}{\ell_\beta^3}
 +\frac{R_{\rm crit}^2}{\ell_G^2}+1=0,
 \label{eq:critical-radius}
\end{equation}
and an equivalent condition is
\begin{equation}
 \frac{R_{\rm crit}^2}{\ell_G^2}=3+\frac{2R_{\rm crit}}{\ell_Q}.
 \label{eq:critical-radius-alternative}
\end{equation}
For continuation in $\beta$ at fixed $(m_\phi,N,a)$, the corresponding coupling is
\begin{equation}
 \beta_{\rm crit}=
 \frac{gN}{2(2\pi)^{3/2}R_{\rm crit}^3}
 \left(1+\frac{R_{\rm crit}}{\ell_Q}
 -\frac{R_{\rm crit}^2}{\ell_G^2}\right).
 \label{eq:critical-beta}
\end{equation}
Equations~\eqref{eq:critical-radius-alternative} and \eqref{eq:critical-beta} locate candidate folds directly. An ordinary fold additionally requires $U'''(R_{\rm crit})\ne0$; the transversality condition for varying $\beta$ follows from $\partial_\beta U'=3N/R_{\rm crit}\ne0$. Degenerate points with vanishing third derivative require a higher-order analysis.

The endpoint behavior clarifies the other sign sectors. At small $R$, the contact term gives $U\to+\infty$ for $g>0$ and $U\to-\infty$ for $g<0$. At large $R$, $U$ grows as $3\beta N\ln R$ for nonzero $\beta$. For simple positive stationary points, see Table~\ref{tab:analytic-sign-sectors}. The ordering follows from the signs of $U'$ at the endpoints and the cubic degree of Eq.~\eqref{eq:equilibrium-length-scales}.
\begin{table}[t]
 \centering
 \begin{tabular}{ccc}
 \hline
 $g$ & $\beta$ & Positive stationary points in increasing $R$\\
 \hline
 $>0$ & $>0$ & one minimum\\
 $>0$ & $<0$ & none, or minimum--maximum\\
 $<0$ & $>0$ & none, or maximum--minimum\\
 $<0$ & $<0$ & one maximum, or maximum--minimum--maximum\\
 \hline
 \end{tabular}
 \caption{Analytically allowed arrangements of simple positive Gaussian equilibria for nonzero couplings. Adjacent extrema can merge at marginal points. A minimum represents stability in the homologous radial sector.}
 \label{tab:analytic-sign-sectors}
\end{table}
For $g>0$, $\beta>0$, finite-energy radial motion is confined between the divergent endpoints of $U$. In sectors with a local minimum and an energetically open endpoint, neighboring maxima define the barriers separating bounded oscillations from large contraction or expansion. For $\beta=0$, $U(\infty)=0$ by convention in Eq.~\eqref{eq:effective-potential}, so the escape threshold can be compared directly with this value.

\begin{figure*}[htbp]
\centering
\includegraphics[width=\linewidth]{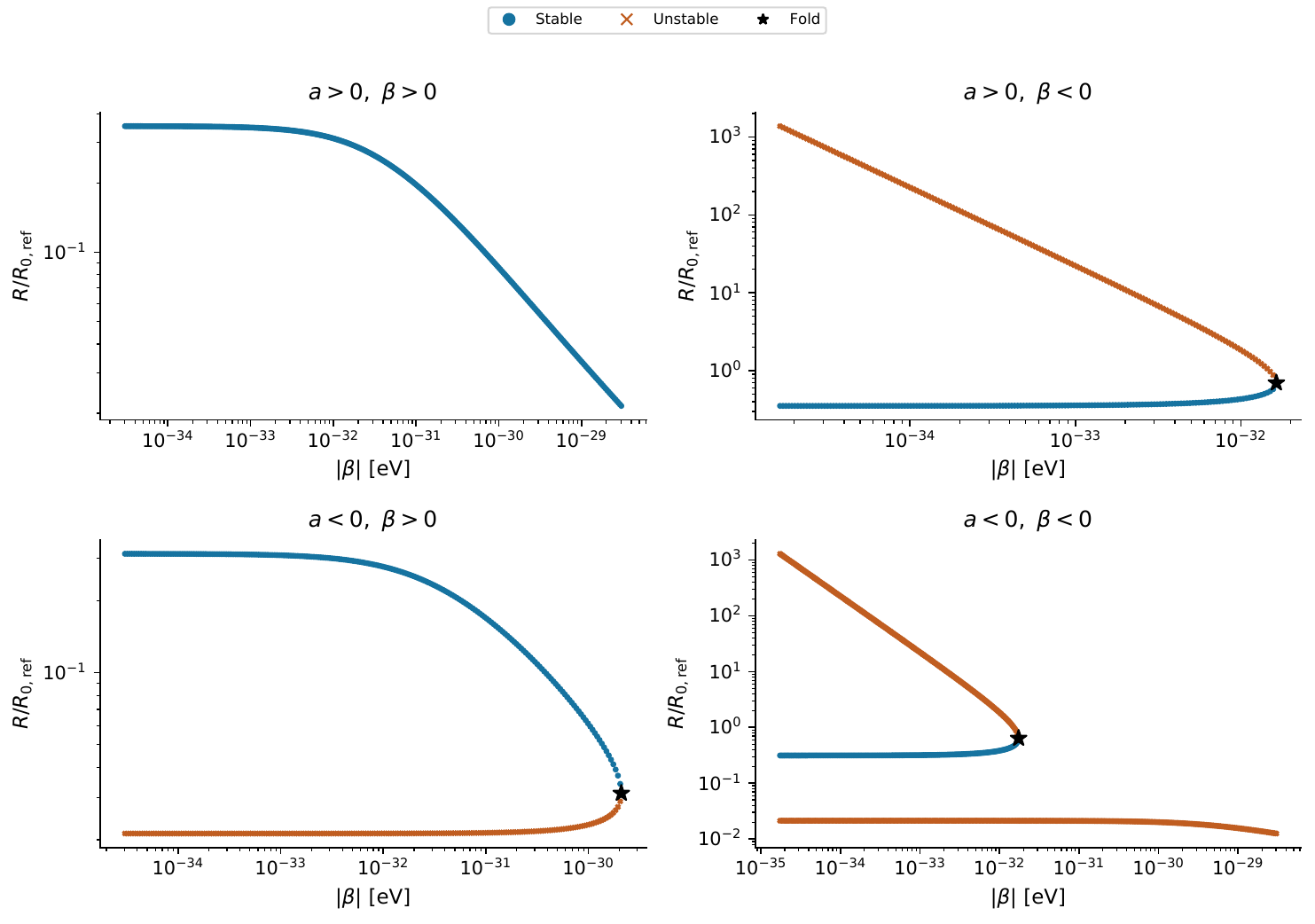}
\caption{Stationary Gaussian radii in the four sign sectors at the fixed parameters stated in the text. Blue circles mark minima, orange crosses maxima, and black stars folds obtained from Eqs.~\eqref{eq:critical-radius-alternative}--\eqref{eq:critical-beta}. Radii are divided by the dwarf-reference width $R_{0,\mathrm{ref}}=2.8260\times10^{20}\,{\rm m}$. The branch experiment uses one hundredth of the positive dwarf reference scattering length. These are stationary branches of the reduced equations; astrophysical admissibility requires separate validity checks.}
\label{fig:signed-branches}
\end{figure*}

Figure~\ref{fig:signed-branches} resolves these arrangements in a controlled continuation at $m_\phi c^2=3.2246\times10^{-23}\,{\rm eV}$, $N=3.4864\times10^{97}$, and $|a|=2.0188\times10^{-79}\,{\rm m}$. For $a>0$, the negative-$\beta$ fold is at $-1.6394\times10^{-32}\,{\rm eV}$. For $a<0$, folds occur at $-1.7401\times10^{-32}\,{\rm eV}$ and $2.0962\times10^{-30}\,{\rm eV}$. Normalized force and curvature residuals are below $10^{-12}$. In the doubly negative sector, the middle stable radius and outer maximum merge at the negative-$\beta$ fold, leaving the inner unstable branch. Thus, the branch diagram identifies how the finite potential well in Fig.~\ref{fig:unified-dynamics} disappears as the coupling changes.

\subsection{Four limiting balances}
For $g>0$ and $\beta>0$, Eq.~\eqref{eq:equilibrium-length-scales} separates the inward contributions $R_0^3/\ell_\beta^3$ and $R_0^2/\ell_G^2$ from the outward contributions $R_0/\ell_Q$ and $1$. Four pairwise balances provide useful asymptotic regimes:
\begin{align}
 \text{Quantum--gravity:}\quad
 &\frac{R_0}{\ell_Q}\simeq\frac{R_0^2}{\ell_G^2},
 & R_0&\simeq\frac{\ell_G^2}{\ell_Q},
 \label{eq:balance-qg}\\
 \text{Contact--gravity:}\quad
 &1\simeq\frac{R_0^2}{\ell_G^2},
 & R_0&\simeq\ell_G,
 \label{eq:balance-cg}\\
 \text{Quantum--logarithmic:}\quad
 &\frac{R_0}{\ell_Q}\simeq\frac{R_0^3}{\ell_\beta^3},
 & R_0&\simeq\left(\frac{\ell_\beta^3}{\ell_Q}\right)^{1/2},
 \label{eq:balance-ql}\\
 \text{Contact--logarithmic:}\quad
 &1\simeq\frac{R_0^3}{\ell_\beta^3},
 & R_0&\simeq\ell_\beta.
 \label{eq:balance-cl}
\end{align}
Each approximation requires both omitted terms to be small compared with the retained balance. Quantum support corresponds to $R_0/\ell_Q\gg1$ and contact support to $R_0/\ell_Q\ll1$. Gravitational confinement dominates when $R_0^2/\ell_G^2\gg R_0^3/\ell_\beta^3$, with the opposite inequality defining logarithmic confinement. At fixed $(m_\phi,a,\beta)$, the four limits give $R_0\propto M^{-1}$, $M^0$, $M^0$, and $M^{1/3}$, respectively. The two constant-radius regimes have different supporting and confining mechanisms. In the contact--logarithmic limit, $\rho_c$ tends to a constant along the sequence.

\subsection{Root search and numerical branch continuation}
The equilibrium analysis classifies the real positive roots of the stationary equation. For numerical evaluation, multiplying Eq.~\eqref{eq:stationarity-dimensional} by $R_0^4/N$ gives the polynomial
\begin{equation}
 3\beta R_0^3+\frac{Gm_\phi^2N}{\sqrt{2\pi}}R_0^2
 -\frac{3\hbar^2}{2m_\phi}R_0
 -\frac{3gN}{2(2\pi)^{3/2}}=0.
 \label{eq:dimensional-polynomial}
\end{equation}
In the branch-continuation calculation, we rescale the radius and coefficients before solving. Vanishing couplings are treated by reducing the polynomial degree explicitly; a vanishing-radius root introduced by multiplication is discarded. We check the branch candidates against the dimensional force balance and the curvature; the archived parameter maps retain the completeness limitations described in Sec.~\ref{sec:results-v5}. For $g\ne0$, a convenient dimensionless equilibrium residual is
\begin{equation}
 \frac{\left|R_0^3/\ell_\beta^3+R_0^2/\ell_G^2-R_0/\ell_Q-1\right|}
 {1+|R_0^3/\ell_\beta^3|+|R_0^2/\ell_G^2|+|R_0/\ell_Q|}.
 \label{eq:root-residual}
\end{equation}
Equation~\eqref{eq:root-residual} tests the equilibrium condition, while Eq.~\eqref{eq:stability-length-scales} determines the radial curvature. Candidate folds satisfy both stationarity and vanishing curvature, as expressed by Eqs.~\eqref{eq:critical-radius-alternative} and~\eqref{eq:critical-beta}, together with the nondegeneracy conditions stated above.

Branch continuation varies $N$ or $\beta$ with the remaining microscopic parameters fixed, retaining stable and unstable roots together. The analytical classification in Sec.~\ref{sec:equilibrium-stability} supplies the expected root structure; scans determine where the corresponding branches lie in the selected parameter domain. Root counts are recorded before and after applying physical cuts, so that a validity boundary is distinguished from a fold. Grid endpoints are reported separately from analytical folds; the archived map boundaries retain the finite resolution of the original sweep.

\section{Results and discussion}
\label{sec:results-v5}

The interaction signs and relative forces organize the equilibrium parameter space. We first examine the radius response and collective pressure, then connect these properties to stellar and halo benchmarks. The parameter slices reveal the associated confinement regimes, while the trial-profile comparison measures the sensitivity of the predicted radii and frequencies.

\subsection{Organization of the equilibrium families}
For repulsive contact interactions and $\beta>0$, the unique Gaussian minimum connects the four limiting balances continuously. The force equalities defined by $\ell_\beta$, $\ell_G$, and $\ell_Q$ identify the transitions between regimes. A direct measure of how the logarithmic interaction changes the radius is obtained by differentiating $U'(R_0)=0$ at fixed $(m_\phi,N,a)$:
\begin{eqnarray}
    \left.\frac{dR_0}{d\beta}\right|_{m_\phi,N,a} &=&-\frac{3N}{R_0U''(R_0)} \\
    \frac{d\ln R_0}{d\ln|\beta|} &=&-\frac{R_0^3/\ell_\beta^3} {1+R_0^2/\ell_G^2+2R_0^3/\ell_\beta^3}.
\label{eq:radius-coupling-response}
\end{eqnarray}
These relations apply along simple stationary branches, with stability specified by $U''(R_0)>0$. Increasing positive $\beta$ contracts a stable cloud. In the quantum--logarithmic and contact--logarithmic limits, the logarithmic slopes approach $-1/2$ and $-1/3$, respectively. The nearly constant radius at sufficiently small $|\beta|$ corresponds to a small numerator in Eq.~\eqref{eq:radius-coupling-response}. The response becomes singular at an ordinary fold as the radial curvature vanishes.

For $a<0$ and $\beta<0$, increasing $|\beta|$ expands a stable configuration. Contact attraction and gravity then compete with gradient support and the expansive logarithmic force. The curvature condition gives the additional restriction
\begin{equation}
    \left|\frac{R_0^2}{\ell_G^2}\right|> 1+2\frac{R_0^3}{\ell_\beta^3}.
    \label{eq:negative-gravity-condition}
\end{equation}
Consequently, gravity exceeds the logarithmic force on every stable Gaussian branch in this sector. This inequality explains the gravitational coloring of the negative-sector maps and connects their stable regions to the potential wells and folds in Figs.~\ref{fig:unified-dynamics} and \ref{fig:signed-branches}.

\subsection{Global variational pressure}
\label{sec:effective-pressure}
For a fixed mass fraction, let $V=4\pi(\kappa R)^3/3$, with $\kappa=\kappa_{99}$ when quoting $R_{99}$. The pressure conjugate to this volume is
\begin{align}
 P(R)&=-\frac{dU/dR}{4\pi\kappa^3R^2}  \\
     &=\frac{3gN^2}{8\pi\kappa^3(2\pi)^{3/2}R^6} \left[1+\frac{R}{\ell_Q}-\frac{R^2}{\ell_G^2} -\frac{R^3}{\ell_\beta^3}\right].
 \label{eq:pressure-R-lengths}
\end{align}
Its volume dependence is obtained directly as
\begin{align}
P(V)={}&
\frac{g\kappa^3N^2}{3\sqrt{2\pi}}V^{-2}
\Biggl[
1+\frac{1}{\ell_Q}
\left(\frac{3V}{4\pi\kappa^3}\right)^{1/3}
\notag\\
&\qquad
-\frac{1}{\ell_G^2}
\left(\frac{3V}{4\pi\kappa^3}\right)^{2/3}
-\frac{1}{\ell_\beta^3}
\frac{3V}{4\pi\kappa^3}
\Biggr].
\label{eq:pressure-V-lengths}
\end{align}
Thus $P(R_0)=0$, and a stable equilibrium has $dP/dR|_{R_0}<0$. This global pressure expresses the mechanical balance of the collective radius. The local interaction pressure follows from the contact and logarithmic energy densities and is $g|\Psi|^4/2-\beta|\Psi|^2$; gradient stress and gravity enter separately in the local hydrodynamic equations \cite{Chavanis:2016ial}.

\subsection{Effective pressure--density relation}
A one-parameter compression sequence at fixed $(m_\phi,N,a,\beta)$ assigns a unique central density $\rho_c=m_\phi N/(\pi^{3/2}R^3)$ to each width. Eliminating $R$ therefore defines an effective equation of state for the collective degree of freedom,
\begin{equation}
P_{\rm eff}(\rho_c;m_\phi,N,a,\beta)
=P\!\left[\left(\frac{m_\phi N}{\pi^{3/2}\rho_c}\right)^{1/3}\right].
\label{eq:effective-eos}
\end{equation}
The quantum, contact, gravitational, and logarithmic contributions scale as $\rho_c^{5/3}$, $\rho_c^2$, $\rho_c^{4/3}$, and $\rho_c$, respectively, with signs fixed by Eq.~\eqref{eq:pressure-R-lengths}. This relation includes the global gravitational binding energy and the gradient energy of the prescribed density profile. Its natural interpretation is the mechanical response of the entire variational cloud at fixed particle number. The associated local material relation follows from $p_{\rm int}=n\,d\mathcal E_{\rm int}/dn-\mathcal E_{\rm int}$:
\begin{equation}
p_{\rm int}(\rho_m)=\frac{g}{2m_\phi^2}\rho_m^2-\frac{\beta}{m_\phi}\rho_m.
\end{equation}
This local interaction equation of state is independent of the reference $\alpha$; gravity and gradient stress enter its hydrodynamic force balance separately. For $\beta<0$, the logarithmic contribution supplies a positive pressure proportional to density, consistent with the isothermal-fluid interpretation of the defocusing logarithmic Schr\"odinger equation \cite{Carles:2021fcy}.
 
\begin{figure}[htbp]
\centering
\includegraphics[width=.990\linewidth]{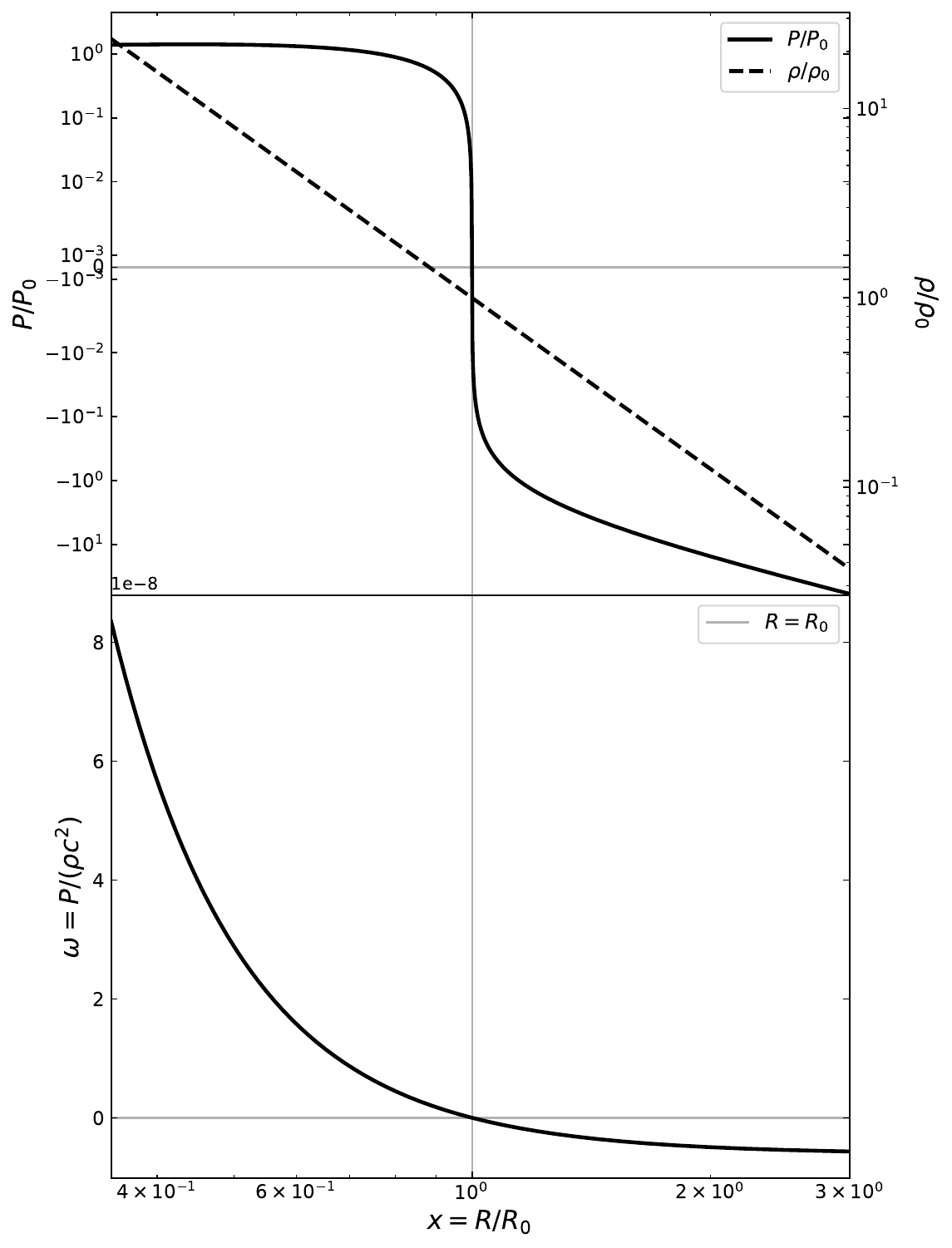}
\caption{Effective pressure and central density along a Gaussian compression sequence. Upper panel: solid $P/P_0$ and dashed $\rho_c/\rho_{c,0}$, with separate vertical axes. The plotted denominator is the radius-dependent contact contribution $P_0=P_{\rm contact}(R)$. Lower panel: $\omega=P/(\rho_c c^2)$. The vertical line marks $R=R_0$. The archived configuration is $m_\phi=4.53\times10^{-60}\,{\rm kg}$, $N=4.45\times10^{98}$, $a=5.71\times10^{-79}\,{\rm m}$, $\beta=2.54\times10^{-50}\,{\rm J}$, and $R_0=2.86\times10^{20}\,{\rm m}$. It is a separate dwarf-scale reference from the grid-selected configurations in Table~\ref{tab:best-micro}.}
\label{fig:pressure-eos}
\end{figure}

Figure~\ref{fig:pressure-eos} shows the collective compression sequence for a positive-sector dwarf-scale configuration. Its horizontal coordinate parametrizes a sequence of cloud widths at fixed particle number and microscopic couplings. The dashed density curve follows $\rho_c/\rho_{c,0}=x^{-3}$. Compression below $R_0$ produces positive collective pressure, while expansion above $R_0$ produces negative pressure and a restoring inward force. At the equilibrium,
\begin{equation}
\left.\frac{dP_{\rm eff}}{d\rho_c}\right|_{R_0}
=\frac{U''(R_0)}{12\pi\kappa_{99}^3R_0\rho_{c,0}}>0,
\end{equation}
which connects the effective compressibility to radial stability. The plotted ratio $\omega=P/(\rho_c c^2)$ measures pressure relative to the central rest-mass energy density. Its zero crossing marks collective mechanical equilibrium; its small magnitude in this example indicates a weak pressure response relative to the rest-energy scale.

\subsection{Benchmarks, selection, and physical cuts}
\label{sec:benchmark-protocol}
We explore 60 logarithmically spaced values along each of the four axes, separately for $(a,\beta)>(0,0)$ and $(a,\beta)<(0,0)$. The sampled magnitudes span $m_\phi c^2=10^{-27}$--$10^{-13}\,{\rm eV}$, $N=10^{75}$--$10^{110}$, $|a|=10^{-95}$--$10^{-75}\,{\rm m}$, and $|\beta|=10^{-80}$--$10^{-30}\,{\rm J}$. Each sector contains $60^4$ parameter combinations. The three targets in Table~\ref{tab:benchmark-definitions} connect the microscopic scan to stellar, dwarf-halo, and cluster-halo scales.

\begin{table}[htbp]\centering
\begin{tabular}{lcccc}\hline
Target & Mass & $R_{\rm target}$ & Convention & $(\delta_M,\delta_R)$\\\hline
Solar & $M_\odot$ & $R_\odot$ & stellar surface & $(.05,.05)$\\
Dwarf & $10^9M_\odot$ & $22\,{\rm kpc}$ & $R_{200}$ & $(.20,.10)$\\
Cluster & $4.1\times10^{14}M_\odot$ & $1\,{\rm Mpc}$ & $R_{500}$ & $(.20,.10)$\\\hline
\end{tabular}
    \caption{Benchmark values and fractional comparison widths. The radius convention describes the target object. The present scan compares the model's $R_{99}$ with $R_{\rm target}$.}
\label{tab:benchmark-definitions}
\end{table}

The comparison uses total model mass $M=m_\phi N$ and its $99\%$ mass radius. An observational overdensity fit imposes $M(<R_\Delta)=(4\pi/3)\Delta\rho_{\rm ref}(z)R_\Delta^3$, specifying the reference density and redshift. The listed $R_{200}$ and $R_{500}$ serve as characteristic target sizes; identifying them with $R_{99}$ is a benchmark convention. The same distinction applies between a Gaussian mass-fraction radius and a stellar surface.

Configurations are ranked by
\begin{equation}
    \chi^2_{\rm MR}=\frac{\ln^2(M/M_{\rm target})}{\ln^2(1+\delta_M)}+
    \frac{\ln^2(R_{99}/R_{\rm target})}{\ln^2(1+\delta_R)}.
    \label{eq:mass-radius-chi-square}
\end{equation}
We use $\chi^2_{\rm MR}$ as a dimensionless discrepancy measure with user-specified fractional comparison widths. The selection $\chi^2_{\rm MR}-\chi^2_{{\rm MR},\min}\leq2.30$ defines a relative neighborhood around each sampled sector minimum. The minimum score and the separate mass and radius ratios characterize proximity to the target. The selected neighborhoods are conditional on this operational window, the grid, and the implemented cuts.

The parameter maps apply the compactness and diluteness cuts
\begin{equation}
    \frac{2GM}{R_{99}c^2}<10^{-2},\qquad
    \frac{N|a|^3}{\pi^{3/2}R_0^3} <\left(\frac{0.05\times15\sqrt\pi}{128}\right)^2 =1.07858\times10^{-4}.
    \label{eq:map-cuts}
\end{equation}
The first tests compactness at $R_{99}$ and the second is the central dilute-gas parameter. Their thresholds define explicit numerical working bounds for the weak-gravity and dilute-gas approximations. For a Gaussian, the central potential ratio in Eq.~\eqref{eq:gravity-dilution-cuts} is $\kappa_{99}/\sqrt\pi\simeq1.344$ times this compactness. Hatching marks stable points failing either implemented cut. The additional kinetic and interaction conditions in Eq.~\eqref{eq:nonrelativistic-cuts} provide a separate audit of the nonrelativistic approximation; the reported counts refer specifically to the two-cut sample in Eq.~\eqref{eq:map-cuts}.

\subsection{Benchmark comparisons across mass scales}
Table~\ref{tab:best-matches} reports all six sector--target comparisons. In the positive sector, the selected radii lie within $1\%$ of their targets, with mass ratios $0.9693$, $1.0079$, and $0.9266$. The corresponding negative-sector optima have substantially larger scores. In particular, the negative solar optimum has $M=4.993M_\odot$ and $R_{99}=6.798R_\odot$, and $2820$ sampled roots lie near that sector minimum. The negative dwarf optimum reproduces the target radius to $0.86\%$ but reaches only $58.4\%$ of the target mass. The negative cluster optimum reaches $121.8\%$ of the target mass and $61.8\%$ of its radius. These outcomes describe this sampled domain and resolution.
\begin{table*}[htbp]\centering\small
\begin{tabular}{llrrrrrr}\hline
Target & Sector & $n_{\rm rel}$ & $\chi^2_{\min}$ & $M/M_t$ & $R_{99}/R_t$ & $C$ & $D$\\\hline
Solar & $+,+$ &63&0.4106&0.9693&0.9976&$4.12\times10^{-6}$&$2.25\times10^{-167}$\\
Dwarf & $+,+$ &112&0.009901&1.0079&0.9915&$4.42\times10^{-9}$&$2.28\times10^{-195}$\\
Cluster & $+,+$ &248&0.1758&0.9266&0.9970&$3.65\times10^{-5}$&$8.66\times10^{-192}$\\
Solar & $-,-$ &2820&2629.405&4.9926&6.7985&$3.12\times10^{-6}$&$1.28\times10^{-174}$\\
Dwarf & $-,-$ &33&8.7325&0.5836&1.0086&$2.52\times10^{-9}$&$2.01\times10^{-197}$\\
Cluster & $-,-$ &25&26.744&1.2177&0.6175&$7.74\times10^{-5}$&$3.79\times10^{-245}$\\\hline
\end{tabular}
\caption{Best scan candidates after the two cuts in Eq.~\eqref{eq:map-cuts}. $n_{\rm rel}$ counts roots within $2.30$ of the sector minimum. $C=2GM/(R_{99}c^2)$ and $D=N|a|^3/(\pi^{3/2}R_0^3)$.}
\label{tab:best-matches}
\end{table*}
\begin{table*}[htbp]\centering\small
\setlength{\tabcolsep}{2.5pt}
\begin{tabular}{llccccc}\hline
Target & Sector & $m_\phi c^2\,[{\rm eV}]$ & $N$ & $|a|\,[{\rm m}]$ & $|\beta|\,[{\rm eV}]$ & Boundary\\\hline
Solar &$+,+$&$3.4864\times10^{-19}$&$3.1012\times10^{84}$&$10^{-75}$&$6.2415\times10^{-12}$&$a,\beta$\\
Dwarf &$+,+$&$3.2246\times10^{-23}$&$3.4864\times10^{97}$&$2.0188\times10^{-77}$&$1.5315\times10^{-58}$&---\\
Cluster &$+,+$&$3.6251\times10^{-24}$&$1.1690\times10^{104}$&$9.6172\times10^{-77}$&$1.4165\times10^{-30}$&---\\
Solar &$-,-$&$10^{-13}$&$5.5688\times10^{79}$&$10^{-75}$&$3.9075\times10^{-44}$&$m_\phi,a$\\
Dwarf &$-,-$&$1.8672\times10^{-23}$&$3.4864\times10^{97}$&$4.2376\times10^{-78}$&$3.0918\times10^{-60}$&---\\
Cluster &$-,-$&$5.1507\times10^{-27}$&$1.0812\times10^{107}$&$10^{-95}$&$4.0630\times10^{-33}$&$a$\\\hline
\end{tabular}
\caption{Microscopic coordinates of the supplied best candidates. The sector gives signs. Boundary entries identify coordinates on the sampled domain edge. Very small logarithmic couplings can give indistinguishable scores at numerical precision; a selected value in this regime is representative of a plateau.}
\label{tab:best-micro}
\end{table*}
The additional energy-scale audit separates mass--radius agreement from nonrelativistic physical consistency. The five halo and negative-solar best candidates satisfy the three equilibrium energy ratios in Eq.~\eqref{eq:nonrelativistic-cuts} to $10^{-2}$; the largest among these ratios is $3.49\times10^{-5}$ for the positive cluster contact term. The positive solar grid optimum requires a different physical interpretation. Its gradient, central contact, and logarithmic ratios in Eq.~\eqref{eq:nonrelativistic-cuts} are approximately $2.83\times10^6$, $9.06\times10^7$, and $1.02\times10^8$. Its accurate mass--radius match therefore lies outside the effective model's nonrelativistic regime. We use a separately refined solar configuration satisfying baseline $10^{-2}$ bounds for the profile comparison below. Each map and table retains its own stated selection, so that the effect of additional validity conditions remains identifiable.

\begin{table*}[htbp]\centering\small
\begin{tabular}{llcc}\hline Target & Quantity & Minimum & Maximum\\\hline
Solar & $m_\phi c^2\,[{\rm eV}]$ & $3.4864\times10^{-19}$ & $8.2272\times10^{-17}$\\
Solar & $N$ & $1.3141\times10^{82}$ & $3.1012\times10^{84}$\\
Solar & $|a|\,[{\rm m}]$ & $10^{-95}$ & $10^{-75}$\\
Solar & $|\beta|\,[{\rm eV}]$ & $2.5437\times10^{-15}$ & $6.2415\times10^{-12}$\\
Dwarf & $m_\phi c^2\,[{\rm eV}]$ & $1.7270\times10^{-27}$ & $3.2246\times10^{-23}$\\
Dwarf & $N$ & $3.4864\times10^{97}$ & $4.9535\times10^{101}$\\
Dwarf & $|a|\,[{\rm m}]$ & $10^{-95}$ & $10^{-75}$\\
Dwarf & $|\beta|\,[{\rm eV}]$ & $6.2415\times10^{-62}$ & $8.5287\times10^{-24}$\\
Cluster & $m_\phi c^2\,[{\rm eV}]$ & $10^{-27}$ & $3.6251\times10^{-24}$\\
Cluster & $N$ & $1.1690\times10^{104}$ & $4.2376\times10^{107}$\\
Cluster & $|a|\,[{\rm m}]$ & $1.4208\times10^{-86}$ & $10^{-75}$\\
Cluster & $|\beta|\,[{\rm eV}]$ & $6.2415\times10^{-62}$ & $2.9734\times10^{-21}$\\
\hline\end{tabular}
\caption{Projected microscopic ranges in the positive-sector relative neighborhoods. Single values denote one sampled coordinate. The ranges are correlated and conditional on the grid and the cuts in Eq.~\eqref{eq:map-cuts}. Couplings are reported as magnitudes.}
\end{table*}

\begin{table}[htbp]\centering\small
\begin{tabular}{llcc}\hline Target & Quantity & Minimum & Maximum\\\hline
Solar & $m_\phi c^2\,[{\rm eV}]$ & $10^{-13}$ & $10^{-13}$\\
Solar & $N$ & $5.5688\times10^{79}$ & $5.5688\times10^{79}$\\
Solar & $|a|\,[{\rm m}]$ & $10^{-95}$ & $10^{-75}$\\
Solar & $|\beta|\,[{\rm eV}]$ & $6.2415\times10^{-62}$ & $6.0026\times10^{-23}$\\
Dwarf & $m_\phi c^2\,[{\rm eV}]$ & $1.8672\times10^{-23}$ & $1.8672\times10^{-23}$\\
Dwarf & $N$ & $3.4864\times10^{97}$ & $3.4864\times10^{97}$\\
Dwarf & $|a|\,[{\rm m}]$ & $4.2376\times10^{-78}$ & $4.2376\times10^{-78}$\\
Dwarf & $|\beta|\,[{\rm eV}]$ & $6.2415\times10^{-62}$ & $8.2023\times10^{-35}$\\
Cluster & $m_\phi c^2\,[{\rm eV}]$ & $5.1507\times10^{-27}$ & $5.1507\times10^{-27}$\\
Cluster & $N$ & $1.0812\times10^{107}$ & $1.0812\times10^{107}$\\
Cluster & $|a|\,[{\rm m}]$ & $10^{-95}$ & $1.3664\times10^{-87}$\\
Cluster & $|\beta|\,[{\rm eV}]$ & $4.0630\times10^{-33}$ & $4.0630\times10^{-33}$\\
\hline\end{tabular}
\caption{Projected microscopic ranges in the negative-sector relative neighborhoods. Single values denote one sampled coordinate. The ranges are correlated and conditional on the grid and the cuts in Eq.~\eqref{eq:map-cuts}. Couplings are reported as magnitudes.}
\end{table}

\subsection{Dwarf-halo parameter space and force balances}
\begin{figure*}[htbp]
\centering
\includegraphics[width=\linewidth]{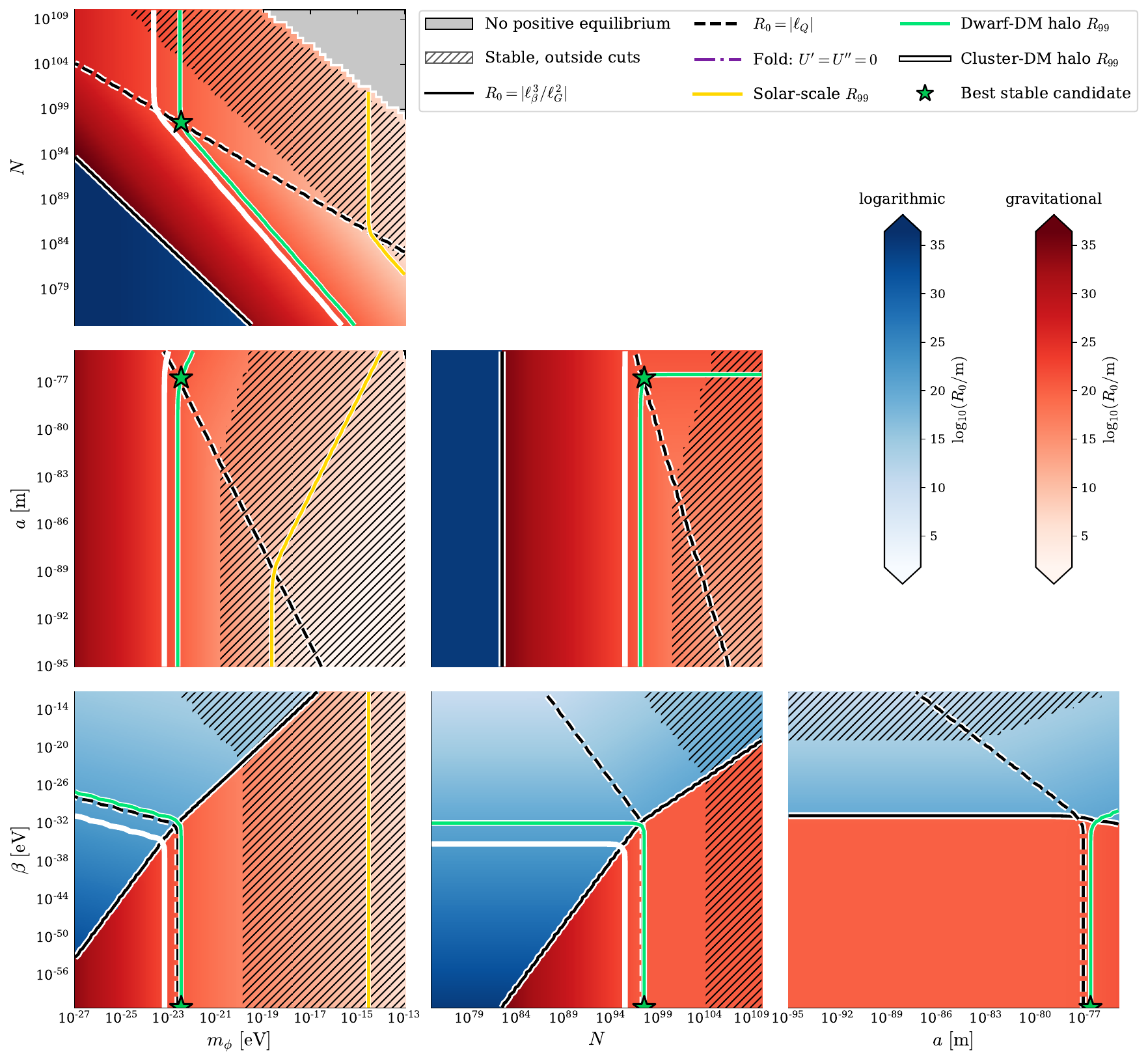}
    \caption{Positive-sector dwarf-halo parameter slices. Stars mark the selected stable reference; two coordinates vary per panel, and the other two are fixed at that reference. Blue denotes stronger logarithmic confinement and red stronger gravitational confinement; intensity gives $\log_{10}(R_0/{\rm m})$. Solid black and dashed curves mark the force equalities specified in the text. Yellow, green, and white contours give the solar, dwarf, and cluster target $R_{99}$, respectively; joint mass--radius selection uses Eq.~\eqref{eq:mass-radius-chi-square}. Hatching uses Eq.~\eqref{eq:map-cuts}. Gray denotes no positive root; analytically, a positive root exists everywhere in this sign sector. The analytical classification fixes the entire positive sector to one minimum; folds occur in the other sign sectors shown in Fig.~\ref{fig:signed-branches}.}
\label{fig:dwarf-positive}
\end{figure*}

\begin{figure*}[htbp]
\centering
\includegraphics[width=\linewidth]{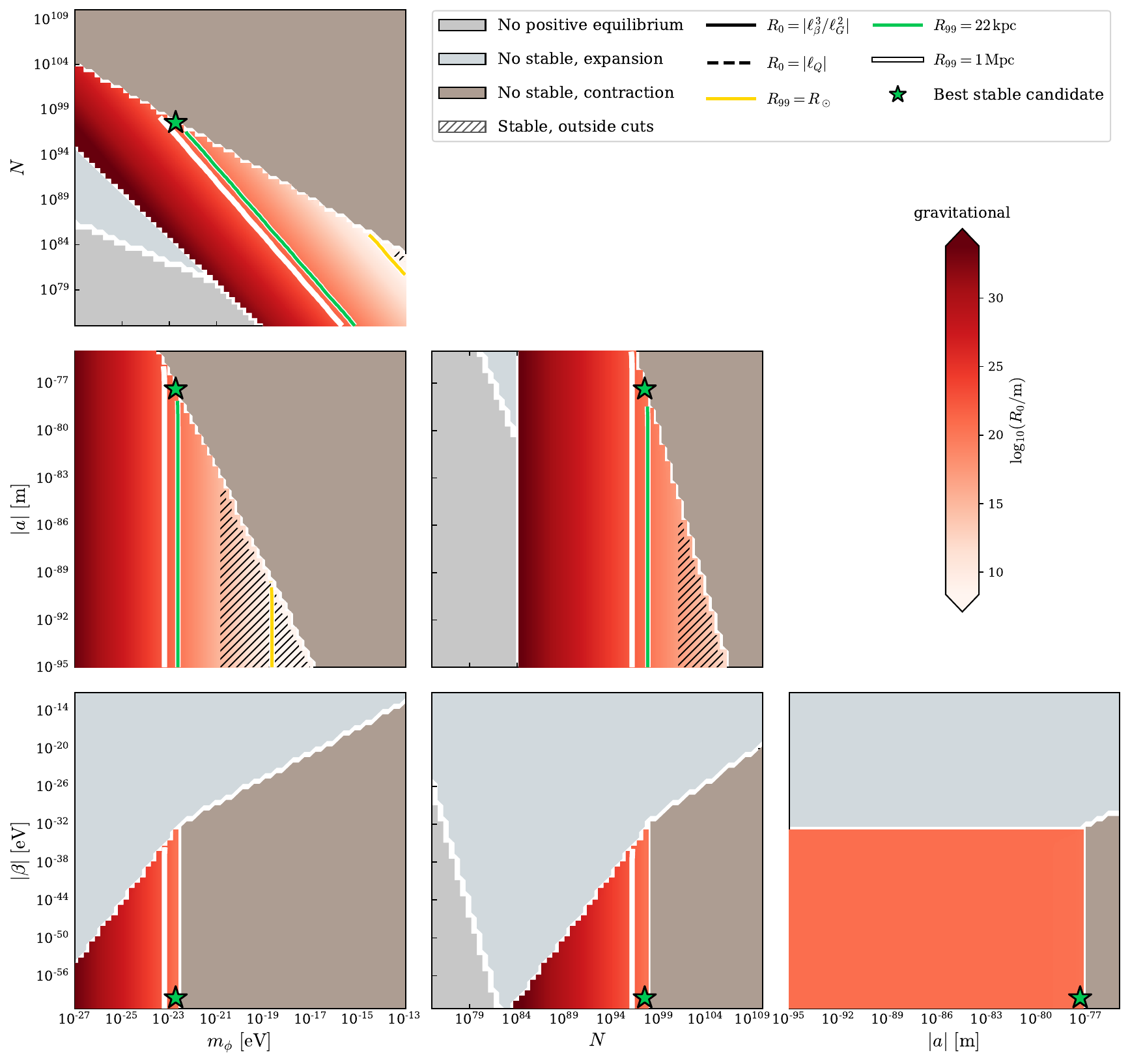}
    \caption{Negative-sector dwarf-halo parameter slices, using $|a|$ and $|\beta|$ on logarithmic axes. Red intensity shows the selected stable radius; Eq.~\eqref{eq:negative-gravity-condition} requires gravitational force dominance on this branch. The other colors record the supplied numerical classifications outside the stable domain. Hatching denotes the compactness or diluteness cut. Stars locate the reference with $M=5.8361\times10^8M_\odot$ and $R_{99}=22.190\,{\rm kpc}$, whose mass--radius score is $8.7325$. The contours and force-comparison curves use the conventions of Fig.~\ref{fig:dwarf-positive}.}
\label{fig:dwarf-negative}
\end{figure*}

Figures~\ref{fig:dwarf-positive} and \ref{fig:dwarf-negative} show six conditional two-parameter slices, with the other coordinates held at the selected reference. They display how equilibrium size and radial stability organize the four-dimensional parameter space. Color intensity represents $\log_{10}(R_0/{\rm m})$. In the positive sector, blue and red distinguish logarithmic and gravitational confinement, separated by
\begin{equation}
\left|\frac{R_0^3}{\ell_\beta^3}\right|=
\left|\frac{R_0^2}{\ell_G^2}\right|,
\qquad R_0=\left|\frac{\ell_\beta^3}{\ell_G^2}\right|.
\end{equation}
The dashed curve $R_0=|\ell_Q|$ compares gradient and contact forces. Together, these boundaries organize the four balances in Sec.~\ref{sec:equilibrium-stability}. They represent changes in force dominance; the positive equilibrium remains a unique minimum throughout this sector.

Near the positive dwarf reference, $R_0/\ell_Q\simeq0.50$ and $R_0^2/\ell_G^2\simeq1.50$, while the logarithmic contribution is negligible. Contact and gradient support therefore balance gravity together. The broad low-$\beta$ plateau makes the radius insensitive to further reductions of $\beta$, explaining the large range of that coupling in the relative neighborhood. Increasing $\beta$ moves the solution toward logarithmic confinement, with nearly vertical or horizontal radius contours in slices where one microscopic coordinate becomes subdominant dynamically. The quoted one-dimensional extrema are projections of correlated accepted points: combinations assembled independently from their extrema need not satisfy the selection.

For the negative-sector slices, $\ell_\beta^3>0$, $\ell_G^2<0$, and $\ell_Q<0$. The gravity-dominance condition in Eq.~\eqref{eq:negative-gravity-condition} explains the exclusively red stable branch in Fig.~\ref{fig:dwarf-negative}. Around the negative dwarf optimum, the ratios are approximately $R_0/\ell_Q=-2.44$ and $R_0^2/\ell_G^2=-1.44$, with negligible logarithmic contribution. Gradient support balances both contact attraction and gravity. As $|\beta|$ grows, the stable band ends when the expanding logarithmic force removes the local well through a fold.

The regions labeled contraction and expansion indicate the force-based classification used by the map outside the retained stable branch. A unique dynamical outcome additionally depends on initial radius, velocity, and barrier energy, as Fig.~\ref{fig:unified-dynamics} demonstrates. The cubic root classification also predicts at least one positive stationary point in the doubly negative sector. Gray areas in the archived maps indicate no positive roots and mark locations that require a completeness check against the analytical classification. The stable domains and their numerical edges should be read with that distinction.

\subsection{Other mass scales and profile sensitivity}
Appendix~\ref{sec:additional-maps} extends the same analysis to solar and cluster targets in both sectors. The positive cluster optimum has $R_0/\ell_Q\simeq1.44\times10^{-6}$, $R_0^2/\ell_G^2\simeq0.94$, and $R_0^3/\ell_\beta^3\simeq0.06$: contact repulsion supplies nearly all outward support, while gravity and the logarithmic term supply approximately $94\%$ and $6\%$ of confinement. The attractive-sector neighborhood is much narrower on the sampled grid, and its best mass--radius score remains larger. The negative solar and cluster references lie on scan boundaries, so their projected parameter ranges still explicitly depend on the explored domain.

The sensitivity of $R_{99}$ to the density tail provides a separate source of variation. Appendix~\ref{sec:profiles} compares Gaussian, exponential, hyperbolic-secant, and super-Gaussian trial densities with identical microscopic parameters within each case, using $\langle r^2\rangle=3R^2/2$ as the common width convention. We minimize each profile independently: the common convention fixes the meaning of $R$, while the equilibrium rms radius can change. Relative to the Gaussian, $R_{99}$ spans $0.933$--$1.441$ for the refined solar case, $0.847$--$1.563$ for the dwarf, and $0.791$--$1.572$ for the cluster. The corresponding frequency ratios span $0.866$--$1$, $0.754$--$1$, and $0.749$--$1.084$.

The exponential tail produces the largest mass-fraction radius, whereas the super-Gaussian gives the smallest. All four profiles retain a stable stationary configuration in the tested cases, while shifts of up to $57\%$ in $R_{99}$ exceed the adopted $10\%$ halo comparison width. Consequently, the force balances and branch topology organize the interpretation, and precise microscopic intervals remain conditional on the trial profile. The comparison quantifies the spread of equilibrium predictions across four specified variational families. In the noninteracting gravitational limit, the Gaussian coefficient $R_{99}=8.96\hbar^2/(GMm_\phi^2)$ differs by about $10\%$ from the numerical Schr\"odinger--Poisson value $9.95$, providing an additional reference for the radius accuracy \cite{Chavanis:2011zi,Chavanis:2011zm}.

\subsection{Physical interpretation}
The combined analysis links changes in microscopic interactions to three observable responses: equilibrium size, the curvature governing radial oscillations, and the pressure generated by collective compression. The low-$|\beta|$ plateaus and the monotonic radius response in Eq.~\eqref{eq:radius-coupling-response} explain the geometry of the parameter slices. The fold conditions identify where a local well terminates, while Eq.~\eqref{eq:negative-gravity-condition} imposes a force restriction throughout the stable doubly negative sector. These analytical relations provide checks on the numerical boundaries and a basis for targeted continuation.

The reported intervals are discrete projections of the sampled equilibrium families. Adjacent grid values differ by factors of $1.73$ in $m_\phi$, $3.92$ in $N$, $2.18$ in $|a|$, and $7.04$ in $|\beta|$. Accordingly, the maps resolve broad force-balance regimes and representative configurations, while narrow benchmark neighborhoods remain conditional on the sampling. The analytical fold and coupling-response relations supply continuous information complementary to these discrete maps.

The four-profile comparison supplies a complementary test of structural sensitivity. Its largest radius shifts motivate direct stationary solutions of Eq.~\eqref{eq:log-gpp} coupled to Poisson's equation at representative force balances and near folds. Linear perturbations about those solutions can assess radial modes beyond the homologous approximation. Together with matched observational radius definitions, these calculations provide a route from the present equilibrium organization to quantitative halo-model tests.

\section{Conclusions}
\label{sec:conclusions}

In this paper, we study the properties of a Newtonian self--gravitating Bose-Einstein condensate using Gross--Pitaevskii theory, where a logarithmic potential describes many-body effects. We have organized the equilibrium structure and conservative radial response of self-gravitating condensates with contact and logarithmic interactions within a Gaussian variational description. The characteristic scales $\ell_\beta$, $\ell_G$, and $\ell_Q$ connect the microscopic couplings to confinement, radial stability, and the geometry of the parameter space. The resulting analysis relates analytical force balances to benchmark mass–radius comparisons across stellar, dwarf-halo, and cluster-halo scales.

For repulsive contact interactions and nonnegative logarithmic coupling, the stationary equation has exactly one positive Gaussian minimum. Quantum or contact support, together with gravitational or logarithmic confinement, produces four limiting balances along this family. By differentiating the stationary condition, we find that increasing positive $\beta$ contracts a stable configuration, with logarithmic radius slopes approaching $-1/2$ and $-1/3$ in the quantum–logarithmic and contact–logarithmic limits. The other sign sectors admit changes in the number of stationary radii through folds. In the doubly negative sector, a stable intermediate radius can lie between two unstable extrema, and its curvature requires $|R_0^2/\ell_G^2|>1+2R_0^3/\ell_\beta^3$. This condition reflects the gravitational-force dominance of the stable negative-sector maps.

The same effective potential determines the oscillation frequency, energy barriers, and collective pressure. Eliminating the Gaussian width at fixed particle number defines an effective pressure–density relation, whose slope at equilibrium is positive precisely when the homologous radial curvature is positive. Conservative trajectories illustrate bounded oscillations and barrier crossing, with their numerical accuracy verified by energy conservation and integration refinement. Physical applications are selected using the weak-gravity, diluteness, and nonrelativistic energy criteria, while the negative-sector trajectory example illustrates the formal barrier structure of the reduced system.

The parameter exploration identifies positive-sector dwarf and cluster candidates with mass ratios $1.01$ and $0.93$ and radius ratios $0.992$ and $0.997$ relative to their targets. Both candidates satisfy the additional equilibrium energy-scale checks. The negative-sector optima locate distinct stable families with larger mass–radius distances on the sampled grid. Relative neighborhoods around each sampled optimum describe correlated microscopic ranges conditional on the grid and implemented selection. At the solar scale, a separately refined positive configuration supplies a simultaneous mass–radius match within the baseline nonrelativistic domain. These comparisons establish representative configurations for further tests with consistently defined observational masses and radii.

Profile sensitivity establishes an important scale for quantitative interpretation: the tested trial densities shift $R_{99}$ by up to approximately $57\%$ and the radial frequency by approximately $25\%$ relative to the Gaussian at fixed microscopic inputs. The analytical branch conditions and coupling response therefore provide the organizing results, while numerical refinement and direct field solutions can sharpen the corresponding parameter intervals. This combination of force balances, stability conditions, and explicit validity tests supplies a framework for connecting logarithmic condensate interactions to the structure and dynamics of self-gravitating bosonic systems.

These configurations provide variational models of logarithmic boson stars in the Newtonian regime: coherent, self-gravitating bosonic condensates whose equilibrium structure combines gradient energy, contact interactions, and the logarithmic contribution $-\beta\ln(\alpha^3|\Psi|^2)$ \cite{visinelli,Liebling:2023review}. The characteristic scales connect their confinement mechanisms and radial response across stellar and halo dimensions.

\clearpage
\appendix
\section{Solar and cluster parameter slices}\label{sec:additional-maps}
The four figures below extend the dwarf analysis using the same grid, force comparisons, and two selection cuts. Tables~\ref{tab:best-matches} and \ref{tab:best-micro}. These are conditional slices, with the two unplotted coordinates fixed. The generic numerical labels and finite-grid caveats described in Sec.~\ref{sec:results-v5} apply here as well.

\begin{figure*}[htbp]
\centering
\includegraphics[width=\linewidth]{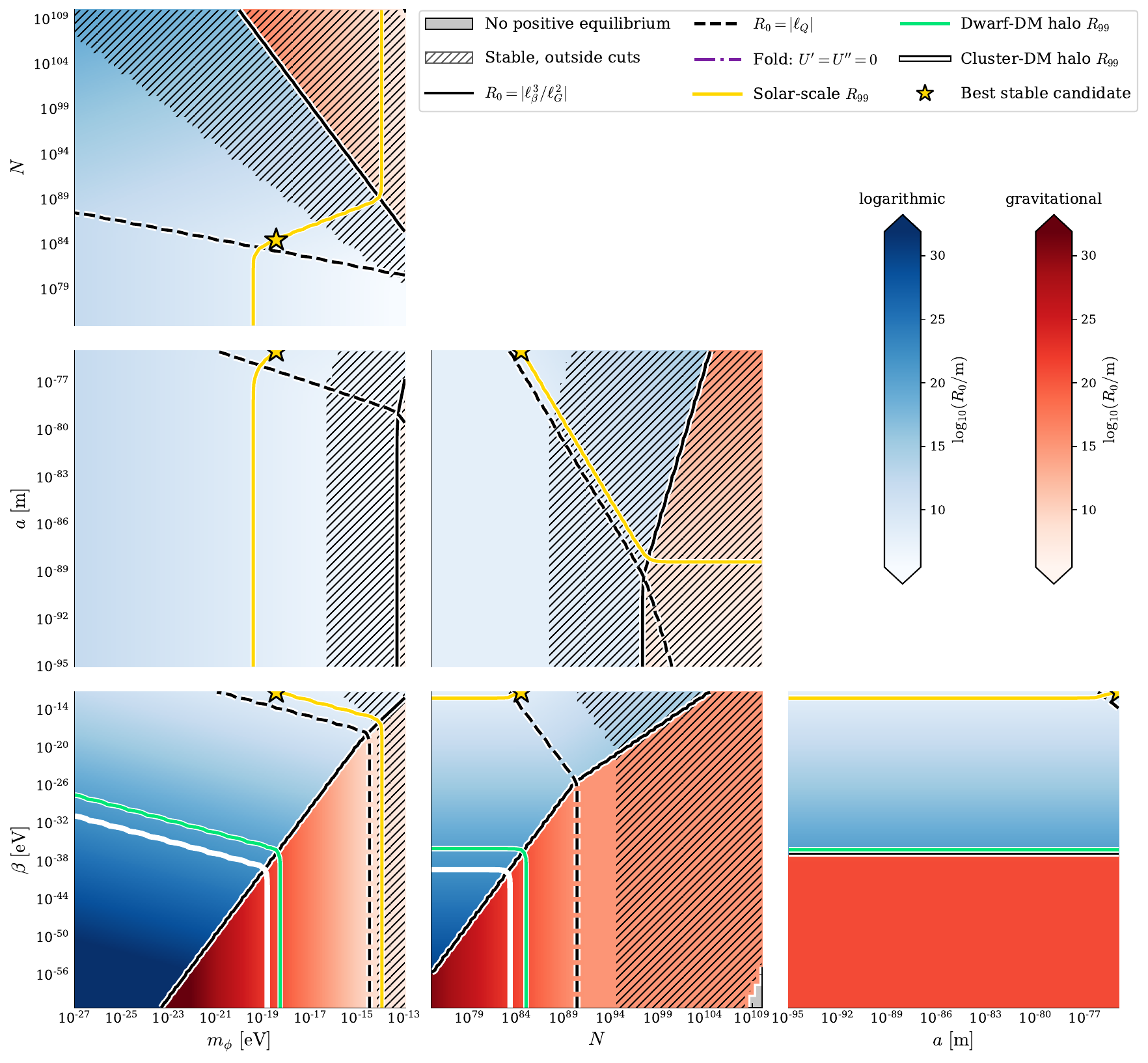}
\caption{Solar-target positive sector. The selected mass and radius closely match the target, but the kinetic and interaction ratios exceed the nonrelativistic domain. Colors, radius contours, and hatching follow Figs.~\ref{fig:dwarf-positive}--\ref{fig:dwarf-negative}.}
\label{fig:solar_scale-positive}
\end{figure*}

\begin{figure*}[htbp]
\centering
\includegraphics[width=\linewidth]{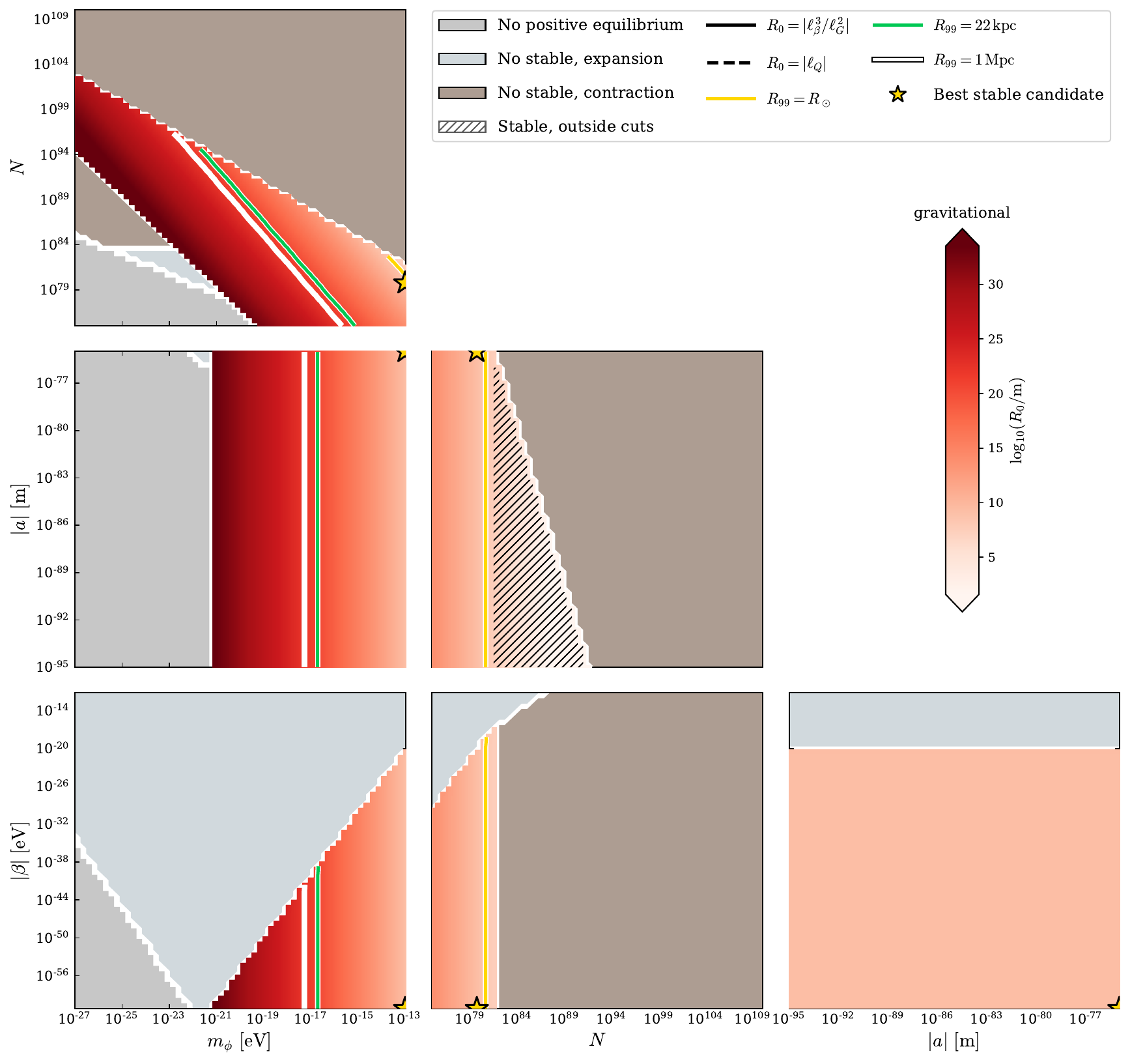}
\caption{Solar-target negative sector. The reference has $M=4.993M_\odot$ and $R_{99}=6.798R_\odot$; it locates the closest sampled stable configuration, with $\chi^2_{\min}=2629.405$. Colors, radius contours, and hatching follow Figs.~\ref{fig:dwarf-positive}--\ref{fig:dwarf-negative}.}
\label{fig:solar_scale-negative}
\end{figure*}

\begin{figure*}[htbp]
\centering
\includegraphics[width=\linewidth]{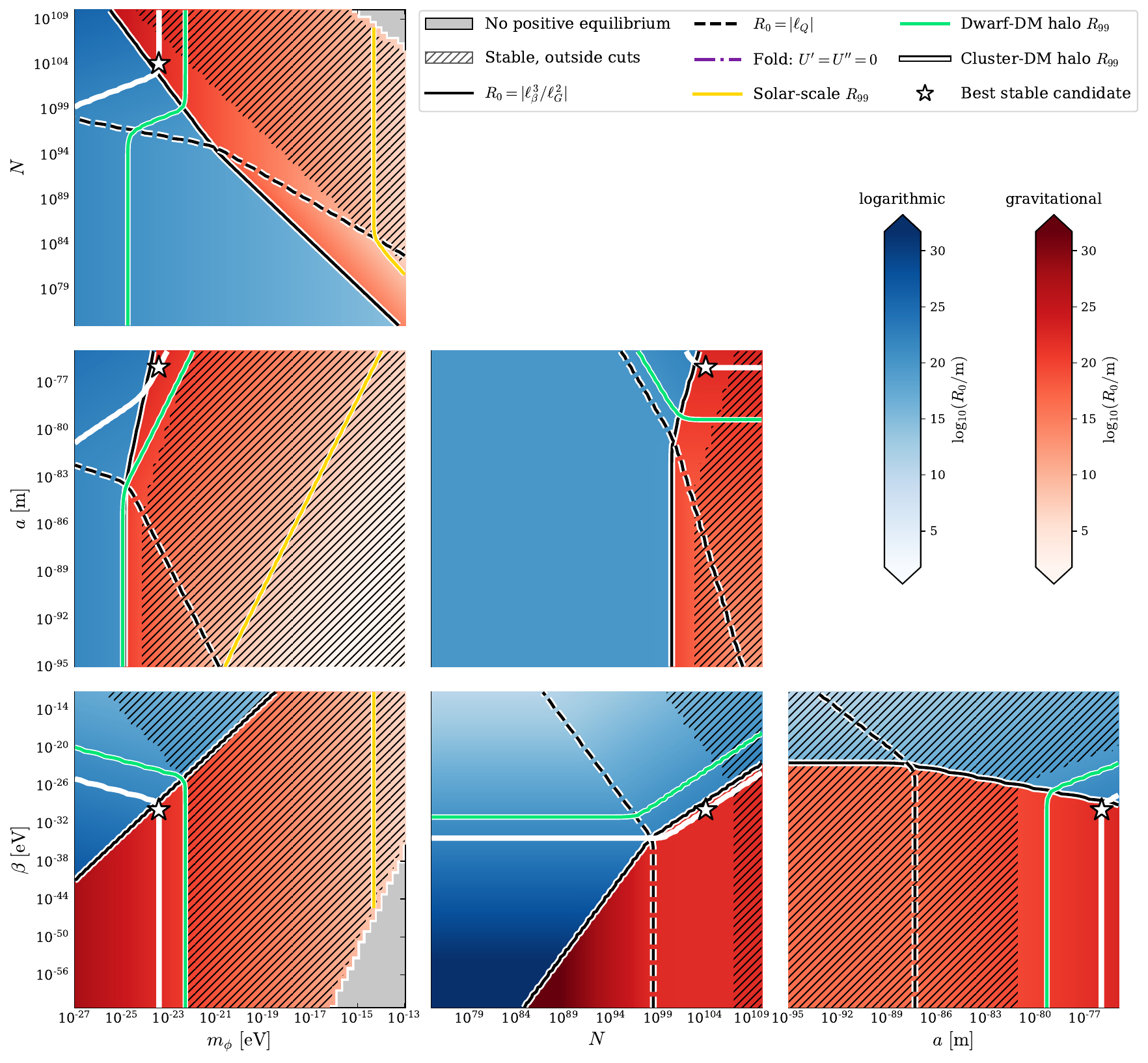}
\caption{Cluster-target positive sector. The reference has $M=3.7991\times10^{14}M_\odot$ and $R_{99}=0.9970\,{\rm Mpc}$. Colors, radius contours, and hatching follow Figs.~\ref{fig:dwarf-positive}--\ref{fig:dwarf-negative}.}
\label{fig:cluster_dm_halo-positive}
\end{figure*}

\begin{figure*}[htbp]
\centering
\includegraphics[width=\linewidth]{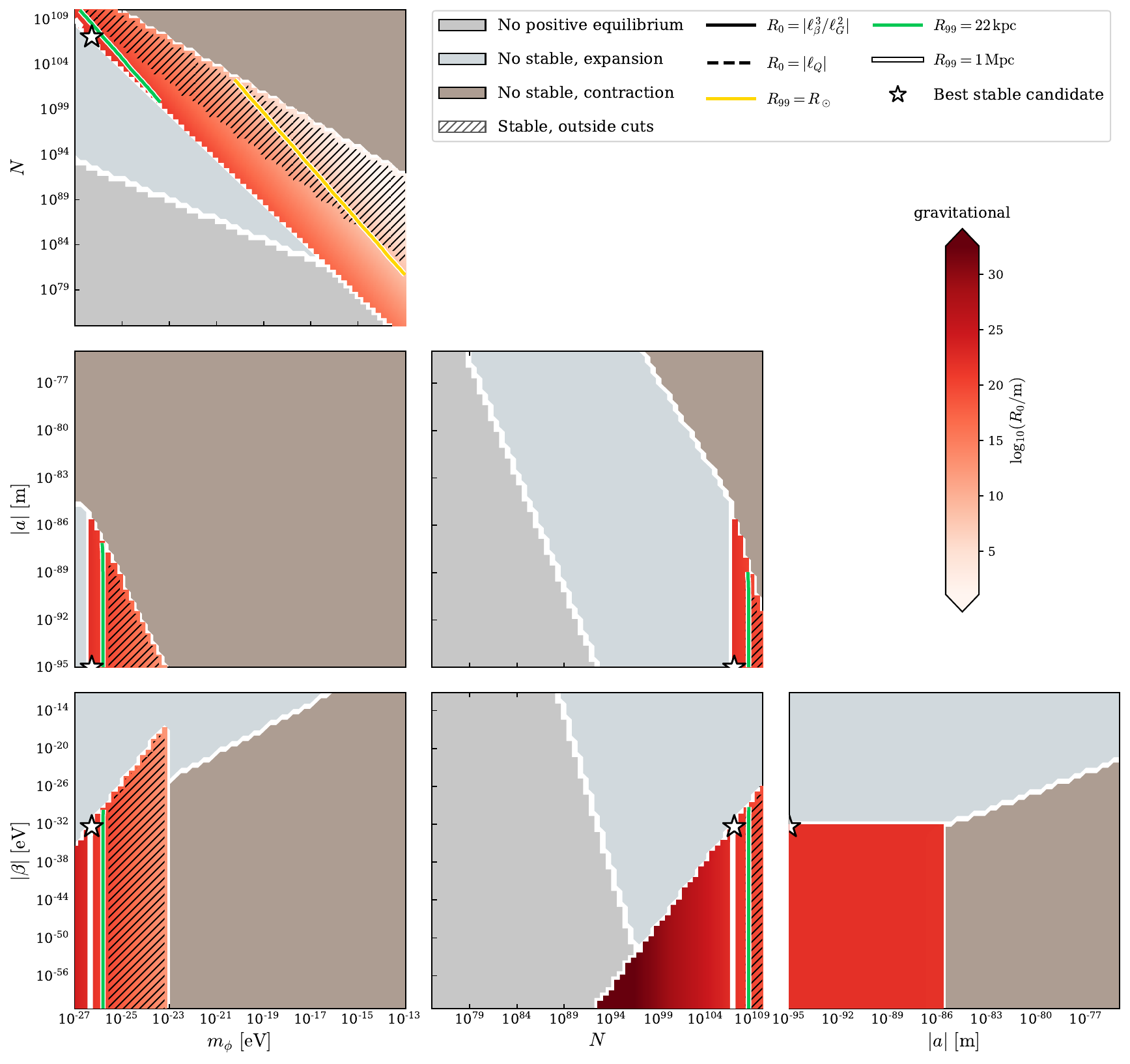}
\caption{Cluster-target negative sector. The reference has $M=4.9926\times10^{14}M_\odot$ and $R_{99}=0.6175\,{\rm Mpc}$ and lies at the lower boundary of $|a|$. Colors, radius contours, and hatching follow Figs.~\ref{fig:dwarf-positive}--\ref{fig:dwarf-negative}.}
\label{fig:cluster_dm_halo-negative}
\end{figure*}

\clearpage
\section{Dependence on the trial density profile}\label{sec:profiles}
\begin{figure}[htbp]
\centering
\includegraphics[width=\linewidth]{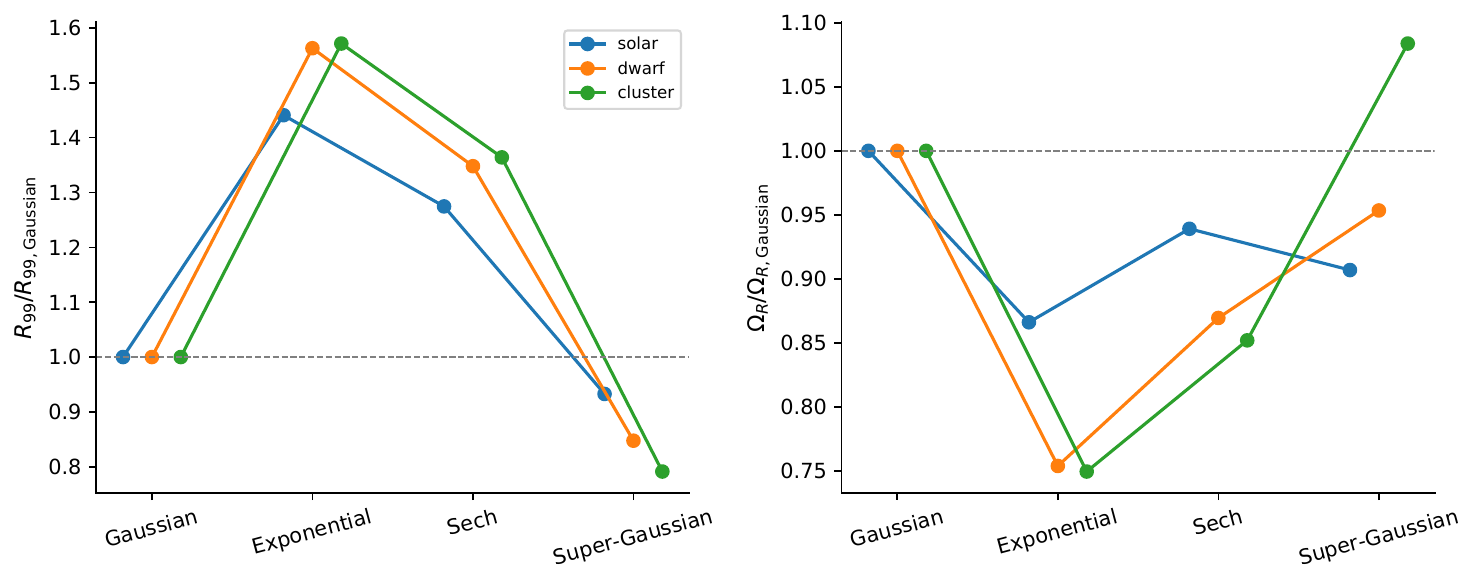}
\caption{Trial-profile dependence at fixed microscopic configurations. Left: equilibrium $R_{99}$ divided by its Gaussian value. Right: radial frequency divided by its Gaussian value. Each profile uses its own enclosed-mass conversion after minimization with the common rms-radius convention. The solar reference is the refined mass--radius match, while dwarf and cluster use the configurations specified in this appendix. Lines connect trial profiles for readability. The ratios quantify profile dependence of the equilibrium predictions.}
\label{fig:profile-dependence}
\end{figure}

Figure~\ref{fig:profile-dependence} uses fixed microscopic configurations for each mass scale, with all four trial densities normalized to $N$ and the same rms-width convention \cite{Eby:2018dat,Castellanos:2020omh}. The refined solar configuration has $m_\phi c^2=1.94149195\times10^{-14}\,{\rm eV}$, $N=5.74514943\times10^{79}$, $a=2.35983347\times10^{-93}\,{\rm m}$, and $\beta=1.17395159\times10^{-17}\,{\rm eV}$. Its Gaussian solution matches $M_\odot$ and $R_\odot$ and is separate from the solar grid optimum in Table~\ref{tab:best-micro}. Its logarithmic validity ratio is $3.43\times10^{-3}$, passing a $10^{-2}$ baseline but exceeding a tightened $10^{-3}$ criterion. The dwarf comparison uses the same $m_\phi,N,a$ as the positive dwarf grid reference with $\beta=6.24150907\times10^{-62}\,{\rm eV}$ on the low-coupling plateau. The cluster uses the positive reference in Table~\ref{tab:best-micro}.

\clearpage
\bibliographystyle{unsrt}
\bibliography{sample}
\end{document}